\documentclass[conference]{IEEEtran}
\usepackage{times}
\usepackage{helvet}
\usepackage{courier}
\usepackage{amsthm}
\usepackage[cmex10]{amsmath}
\usepackage{mathrsfs}
\usepackage{subfigure}
\usepackage{graphicx}
\usepackage{cite}
\usepackage[lined,boxed,ruled]{algorithm2e}
\usepackage{mathrsfs,amsfonts,amsmath,amsthm}
\usepackage{multirow}

\usepackage{kantlipsum} 

\DeclareMathOperator*{\argmax}{arg\,max}
\usepackage[justification=centering]{caption}
\usepackage{color}

\usepackage{paralist}
\newtheorem{definition}{Definition}

\begin{document}

\title{GT-SEER: Geo-Temporal SEquential Embedding Rank for Point-of-interest Recommendation}

\author{Shenglin Zhao, Tong Zhao, Irwin King, and Michael R.~Lyu\\
Department of Computer Science \& Engineering\\ The Chinese University of Hong Kong,
Shatin, N.T., Hong Kong\\
}

\maketitle

\begin{abstract}
Point-of-interest (POI) recommendation is an important application in location-based social networks (LBSNs), which learns the user preference and mobility pattern from check-in sequences to recommend POIs. However, previous POI recommendation systems model check-in sequences based on either tensor factorization or Markov chain model, which cannot capture contextual check-in information in sequences. The contextual check-in information implies the complementary functions among POIs that compose an individual's daily check-in sequence. In this paper, we exploit the embedding learning technique to capture the contextual check-in information and further propose the  \textit{{\textbf{SE}}}quential \textit{{\textbf{E}}}mbedding \textit{{\textbf{R}}}ank (\textit{SEER}) model for POI recommendation. In particular, the \textit{SEER} model learns user preferences via a pairwise ranking model under the sequential constraint modeled by the POI embedding learning method. Furthermore, we incorporate two important factors, i.e.,  temporal influence and geographical influence, into the \textit{SEER} model to enhance the POI recommendation system.  
Due to the temporal variance of sequences on different days, we propose a temporal POI embedding model and incorporate the temporal POI representations into a temporal preference ranking model to establish the \textit{T}emporal \textit{SEER} (\textit{T-SEER}) model. In addition,  We incorporate the geographical influence into the  \textit{T-SEER} model
 and develop the \textit{\textbf{Geo-Temporal}} \textit{{\textbf{SEER}}} (\textit{GT-SEER}) model.
To verify the effectiveness of our proposed methods, we conduct elaborated experiments on two real life datasets. 
Experimental results show that our proposed methods outperform state-of-the-art models. Compared with the best baseline competitor, the \textit{GT-SEER} model improves at least 28\% on both datasets for all metrics.      
\end{abstract}

\section{Introduction}
\label{sec:intro}
Location-based social networks (LBSNs) such as Foursquare have become popular services to attract users sharing their check-in behaviors, making friends, and writing comments on point-of-interests (POIs). 
For example, Foursquare has attracted over 50 million people worldwide and recorded over 8 billion check-ins until now.\footnote{https://foursquare.com/about} 
To improve user experience in LBSNs by suggesting favorite locations, POI recommendation comes out, which mines users' check-in sequences to recommend  places where an individual has not been. POI recommendation  not only helps users explore new interesting places in a city, but also facilitates business owners  to launch advertisements. 
Due to the significance of POI recommendation, a bunch of methods have been proposed to enhance the POI recommendation system~\cite{cheng2012fused,gao2013exploring,DBLP:conf/icdm/LiHZG15,ye2011exploiting,yin2013lcars}.

In general, researchers learn the user preference and the sequence information to recommend POIs~\cite{liu2013personalized,ye2013s,zhang2014lore}. The collaborative filtering techniques are used to learn the user preference~\cite{cheng2013you,DBLP:conf/icdm/LiHZG15,Li:2015:RRB:2766462.2767722,DBLP:conf/icdm/LianGZYXZR15,ye2011exploiting}.
In addition, the tensor factorization and Markov chain model are employed to capture the check-ins' sequential pattern. 
For instance, researchers in~\cite{liu2013personalized,ye2013s} exploit the categories' transitive pattern in sequential check-ins to recommend  POIs. 
Zhang et al.~\cite{zhang2014lore} propose an additive Markov chain model to explore the whole past sequence's influence. 
Moreover,  researchers in~\cite{cheng2013you,feng2015personalized} learn two successive check-ins' transitive probability in latent feature space via a tensor factorization model to recommend next new POIs. 
Although all previous studies have improved POI recommendation from the sequential modeling perspective, they cannot capture contextual check-in information from the whole sequence.

In fact, POIs within a check-in sequence that traces an individual's daily activities always demonstrate a contextual and complementary property.
For example, users always check-in at restaurant, gym, and office within the same sequence of one day. The three types of POIs compose a user's daily life---dining, work, and entertainment after work. Hence, POIs in a sequence are complementary from the function perspective and are highly correlated with such a contextual property. These facts motivate us to come up with an embedding method to capture the contextual information. 

We exploit the embedding learning technique to capture the contextual check-in information and further propose the  \textit{{\textbf{SE}}}quential \textit{{\textbf{E}}}mbedding \textit{{\textbf{R}}}ank (\textit{SEER}) model for POI recommendation.  Specifically, we learn the POI embeddings based on a popular neural language model, \textit{word2vec}~\cite{mikolov2013distributed}. We treat each user as a ``document", check-in sequence in one day as a ``sentence", and each POI as a ``word". Then, we learn the POI representation from check-in sequences in the embedding space. 
On the other hand, we treat the check-in activity as a kind of feedback and learn user preferences through a pairwise ranking model.
In other words, we assume that a user prefers a checked-in POI than the unchecked, and learn this kind of pairwise preference via a ranking model.
On basis of the POI embedding model and the pairwise preference ranking model, we propose the \textit{SEER} model to combine them together.

Moreover, we incorporate two important factors, i.e.,  temporal influence and geographical influence, into the \textit{SEER} model to enhance system performance and propose the \textit{\textbf{T}}emporal \textit{\textbf{SEER}} (\textit{T-SEER}) model and the \textit{\textbf{G}}eo-\textit{\textbf{T}}emporal \textit{\textbf{SEER}} (\textit{GT-SEER}) model.
Because user check-ins in LBSNs are time-sensitive, sequences on different days exhibit temporal variance. For example, users always check-in at POIs around offices on \texttt{weekday} while visit shopping malls on \texttt{weekend}. Therefore, check-in sequences on different days naturally exhibit variant temporal characteristics, ``work'' on \texttt{weekday} and ``entertainment'' on \texttt{weekend}.
To this end, we define the temporal POI, which refers to a POI taking a specific temporal state (i.e., day type, \texttt{weekday} or \texttt{weekend}) as context. Then, we learn the temporal POI embedding given the concatenation of the context POI and the temporal state. 
We incorporate the temporal POI embeddings into a temporal preference ranking model to establish \textit{T-SEER} model.
In addition, we observe that users prefer to visit POIs that are geographically adjacent to their checked-in POIs. 
This geographical characteristic inspires us to advance the preference ranking model through more sophisticated pairwise preference relations that discriminate the unchecked POIs according to geographical information.
Hence, we incorporate the geographical influence into the  \textit{T-SEER} model
 and develop the  \textit{GT-SEER} model.

The contributions of this paper are summarized as follows:
\begin{itemize}
\item By projecting every POI into one object in an embedding space, we learn POIs' contextual relations from check-in sequences through \textit{word2vec} framework. Our proposed  \textit{SEER} model better captures the sequential pattern, learning not only the consecutive check-ins' transitive probability but also POIs' intrinsic relations represented in sequences.  Compared with previous sequential model, the \textit{SEER} model achieves more than 50\% improvement. 
\item We propose the \textit{T-SEER} model that is the first work capturing the variant temporal features in sequences on different days. In addition, our model jointly learns the user preference and sequence pattern. By incorporating the temporal influence, the \textit{T-SEER} model improves the  \textit{SEER} model about 10\%.
\item By exploiting a new way to incorporate the geographical influence, we develop the \textit{GT-SEER} model that improves the \textit{T-SEER} model about 15\%. From the model perspective, we advance the pairwise preference ranking method through discriminating the unchecked POIs according to geographical information. 
\end{itemize}

The rest of this paper is organized as follows. In Section~\ref{sec:rw}, we review the related work. In Section~\ref{sec:dda}, we introduce two real world datasets and report empirical data analysis  that motivates our methods. 
Next, we introduce our proposed mothods, \textit{SEER}, \textit{T-SEER}, and \textit{GT-SEER} model in Section~\ref{sec:model}. Then, we evaluate our proposed models in Section~\ref{sec:ee}. Finally, we conclude this paper and point out possible future work in Section~\ref{sec:cfw}. 

\section{Related Work}
\label{sec:rw}
In this section, we first demonstrate the recent progress of POI recommendation. Then, we report how the prior work exploits the sequential influence, temporal influence, and geographical influence to improve the POI recommendation. Since our proposed methods adopt an embedding learning method, \textit{word2vec}, to model check-in sequences, we also review the literature of \textit{word2vec} framework and its applications.

\textbf{POI Recommendation.} POI recommendation has attracted intensive academic attention recently. Most of proposed methods base on the Collaborative Filtering (CF) techniques to learn user preferences on POIs. Researchers in~\cite{ye2011exploiting,yuan2013time,Zhang:2013:IPG:2525314.2525339} employ the user-based CF to recommend POIs, while,  other researchers~\cite{cheng2012fused,gao2013exploring,gao2015content,DBLP:conf/icdm/LiHZG15,DBLP:conf/icdm/LianGZYXZR15} leverage the model-based CF, i.e., Matrix Factorization (MF)~\cite{koren2009matrix}. Furthermore, Some researchers~\cite{lian2014geomf,liu2014exploiting} observe that it is better to treat the check-ins as implicit feedback than the explicit way. They utilize the weighted regularized MF~\cite{hu2008collaborative} to model this kind of implicit feedback. Other researchers model the implicit feedback through the pairwise learning techniques, which assume  users prefer the checked-ins POIs than the unchecked. Researchers in~\cite{cheng2013you, zhao2016stellar} learn the pairwise preference via the Bayesian personalized ranking (BPR) loss~\cite{rendle2009bpr}.  Li et al.~\cite{Li:2015:RRB:2766462.2767722} propose a ranking based CF model to recommend POIs, which measures the pairwise preference through the WARP loss~\cite{weston2012latenticml}. 

\textbf{Sequential Influence.} Sequential influence is mined for POI recommendation. Existing studies employ the Markov chain property in consecutive check-ins to capture the sequential pattern. Specifically most of successive POI recommendation systems depend on the sequential correlations in successive check-ins~\cite{cheng2013you,feng2015personalized,liu2016predicting,Zhang:2015:LTA:2806416.2806564}.
Researchers in~\cite{cheng2013you,feng2015personalized} recommend the successive POIs on the basis of Factorized Personalized Markov Chain (FPMC) model~\cite{rendle2010factorizing}. 
Liu et al.~\cite{liu2016predicting} employ the recurrent neural network (RNN) to find the sequential correlations. 
In addition, researchers in~\cite{liu2013personalized,ye2013s} learn the categories' transitive pattern in sequential check-ins.
Zhang et al.~\cite{zhang2014lore} predict the sequential transitive probability through an additive  Markov chain model. 
However, all previous sequential models cannot capture contextual check-in information from the whole sequence. 
Hence, we propose a POI embedding method to learn sequential POIs' representations, which captures the check-ins' contextual relations in a sequence. 

\textbf{Temporal Influence.} 
Temporal influence is mined for POI recommendation in prior work~\cite{cheng2013you,cho2011friendship,gao2013exploring,yuan2013time}. 
Temporal characteristics can be summarized as, periodicity, non-uniformness, and consecutiveness. 
Periodicity is first proposed in~\cite{cho2011friendship}, depicting the periodic pattern of user check-in activities. For instance, people always stay in their offices and surrounding places on weekdays while go to shopping malls on weekends. 
Non-uniformness is first proposed in~\cite{gao2013exploring}, demonstrating that a user's check-in preferences may change at different time. 
For example, \texttt{weekday} and \texttt{weekend} imply different check-in preferences, ``work" and ``entertainment". 
In addition, consecutiveness are used in~\cite{cheng2013you,gao2013exploring}, capturing the consecutive check-ins' correlations to improve performance. 
In our model, the consecutiveness can  be depicted in sequential modeling. Moreover, we propose the temporal POI embedding model to capture the periodicity and non-uniformness among \texttt{weekday} and \texttt{weekend}.

\textbf{Geographical Influence.} Geographical influence plays an important role in POI recommendation, since the check-in activity in LBSNs is limited to geographical conditions.    
To capture the geographical influence, researchers in~\cite{cheng2012fused,cho2011friendship,zhao2013capturing} propose Gaussian distribution based models. Researchers in~\cite{ye2011exploiting,yuan2013time} employ the power law distribution model. In addition, researchers in~\cite{Zhang:2013:IPG:2525314.2525339,Zhang:2015:GEG:2766462.2767711,zhang2015orec} leverage the kernel density estimation model. Moreover, researchers in~\cite{lian2014geomf,liu2014exploiting} incorporate the geographical influence into a weighted regularized MF model~\cite{hu2008collaborative, pan2008one} and learn the geographical influence jointly with the user preference. Similar to~\cite{lian2014geomf,liu2014exploiting}, we model the check-ins as implicit feedback; yet we learn it through a Bayesian pairwise ranking method~\cite{rendle2009bpr}. Furthermore, we propose a geographical pairwise ranking model, which captures the geographical influence via discriminating the unchecked POIs according to their geographical information.

\textbf{Embedding Learning.} \textit{Word2vec}~\cite{mikolov2013distributed} is an effective method to learn embedding representations in word sequences.  It models the words' contextual correlations in word sentences, showing better performance than the perspectives of word transitivity in sentences and word similarity. It is generally used in natural language processing~\cite{mikolov2013exploiting,mikolov2013linguistic}. Afterwards, paragraph2vector~\cite{le2014distributed} and other variants~\cite{liu2015learning,liu2015topical} are proposed to enhance the \textit{word2vec} framework for specific purposes. 
 Since the efficacy of the framework in capturing the correlations of items, \textit{word2vec} is employed to the network embedding~\cite{perozzi2014deepwalk},  user modeling~\cite{tang2015user},
 as well as in item modeling~\cite{tang2015learning} and item recommendation~\cite{Grbovic2015E,ozsoy2016word}. These successes persuade us to exploit the \textit{word2vec} framework to model POIs' representations in check-in sequences. Our POI embedding model is similar to the prod2vec model in~\cite{Grbovic2015E} and KNI model in~\cite{ozsoy2016word}. However, we incorporate the temporal variance into the \textit{word2vec} framework to develop the temporal POI embedding that is a variant matching the POI recommendation task.  
\begin{table}[!t]
\centering
\caption{Data statistics}
\begin{tabular}{l|c|c}
\hline
 & \textbf{Foursquare} & \textbf{Gowalla}\\ \hline
\#users & 10,034 & 3,240\\ \hline
\#POIs &  16,561&33,578\\ \hline
\#check-ins & 865,647&556,453\\ \hline
Avg. \#check-ins of each user & 86.3 &171.7\\ \hline
Density & 0.0015 &0.0028\\ \hline
\end{tabular}
\label{tbl:fstat}

\end{table}

 \begin{figure}[t!]%
\centering
 \subfigure[Sequence vs. Random]{
 \includegraphics[height=1.4in,width=1.5in]{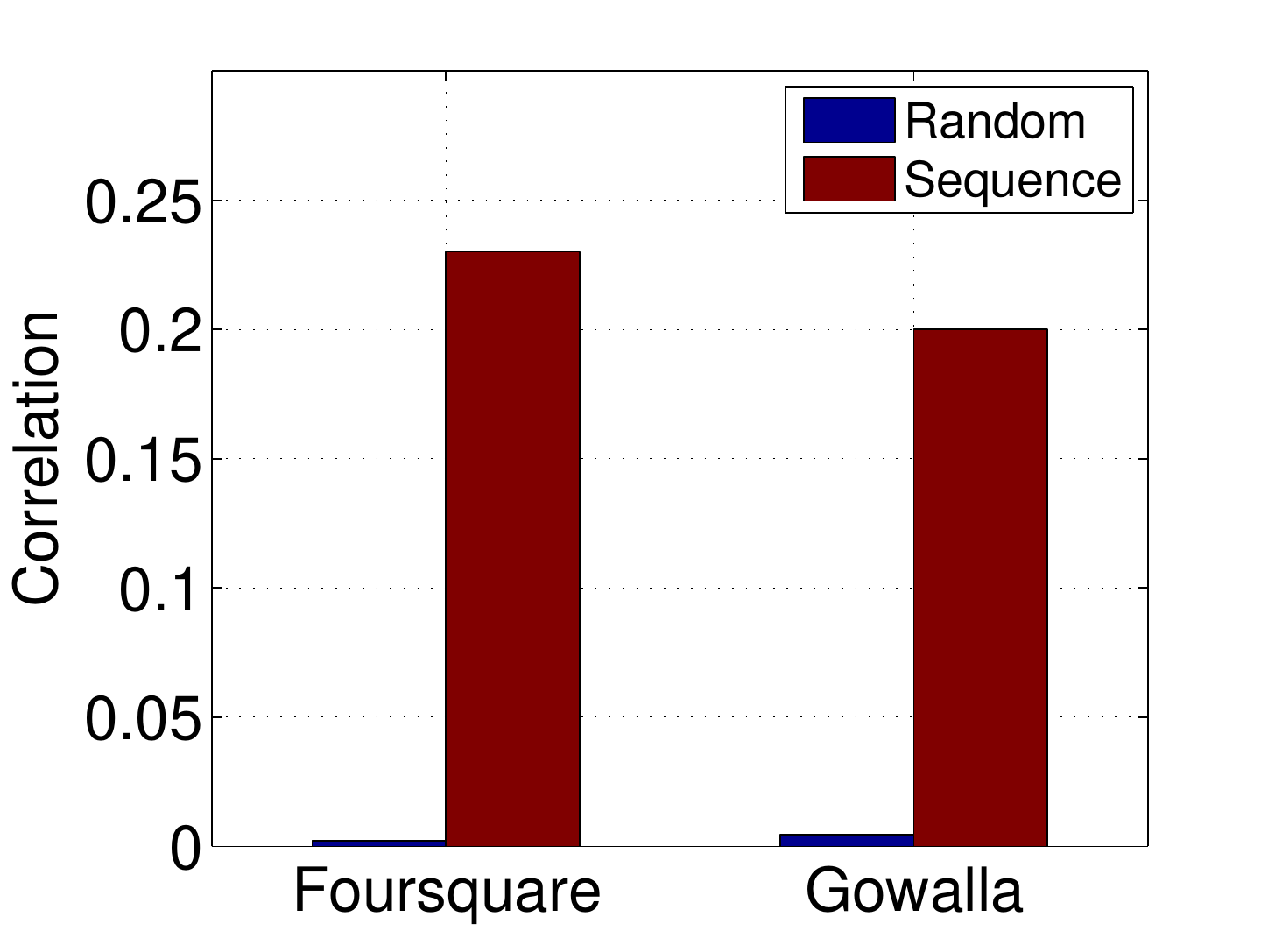}
 \label{subfig:interpoi} }
 \subfigure[\small{Consecutive vs. Nonconsecutive}]{
 \includegraphics[height=1.4in,width=1.6in]{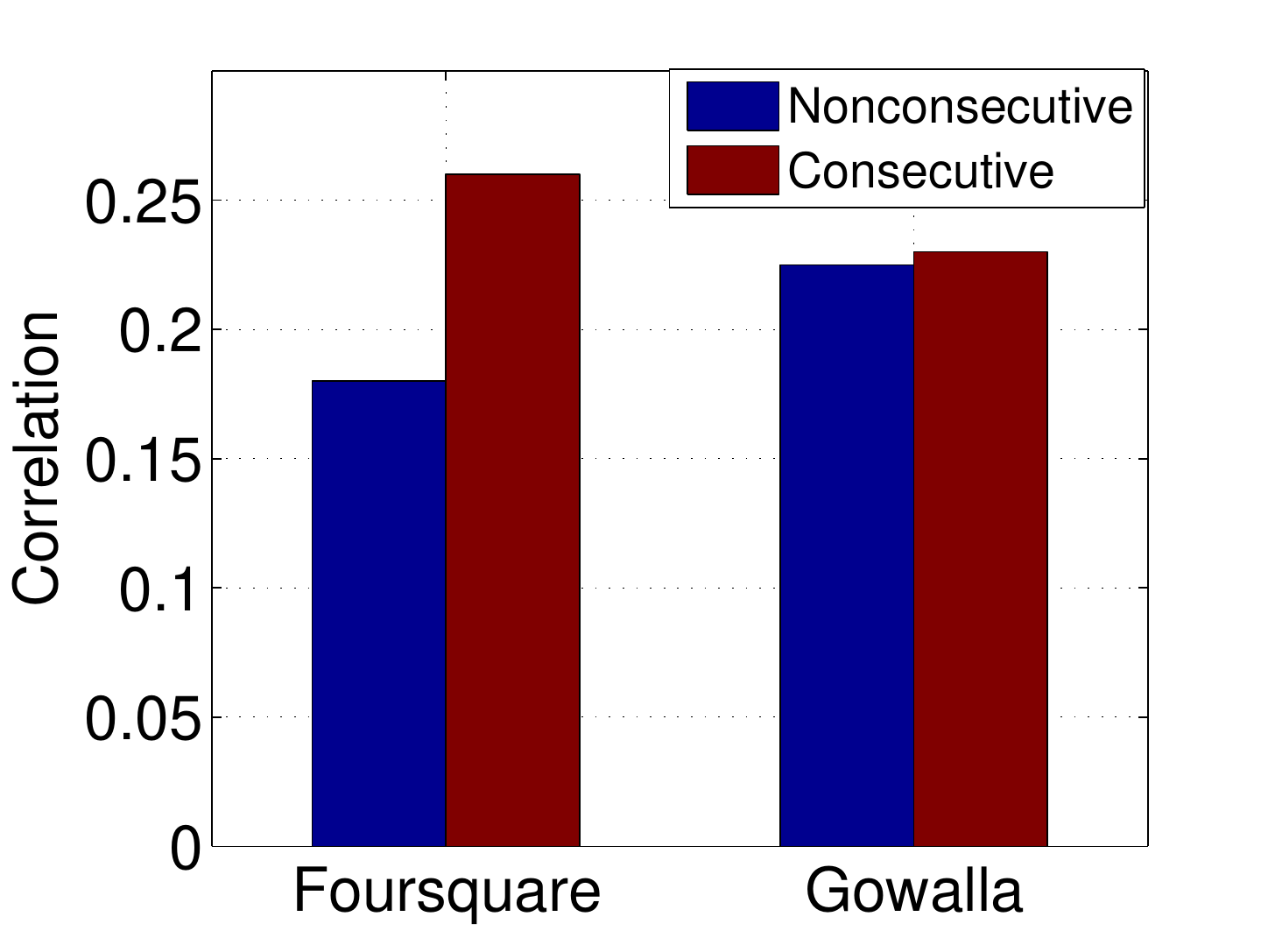}
 \label{subfig:interpoig} }
 \caption{POI correlation in sequences}
 \label{fig:ccdf}

 \end{figure} 

\section{Data Description and Analysis}   
\label{sec:dda}
In this section, we first introduce two real world LBSN datasets,  and then conduct empirical analysis on them to explore the  properties in check-in sequences of one day.
\subsection{Data Description}
We use two check-in datasets crawled from real world LBSNs: Foursquare data  provided in~\cite{gao2012gscorr} and Gowalla data in~\cite{zhao2013capturing}.  We preprocess the data by filtering the POIs checked-in  less than five users and users whose check-ins are less than ten times. Then we keep the remaining users' check-in records from January 1, 2011 to July 31, 2011. After the preprocessing, the datasets contain the statistical properties as shown in Table~\ref{tbl:fstat}.

\begin{figure}[t!]
\centering
\subfigure[Foursquare]{\label{fig:4s_month}\includegraphics[height=1.8in,width=\columnwidth]{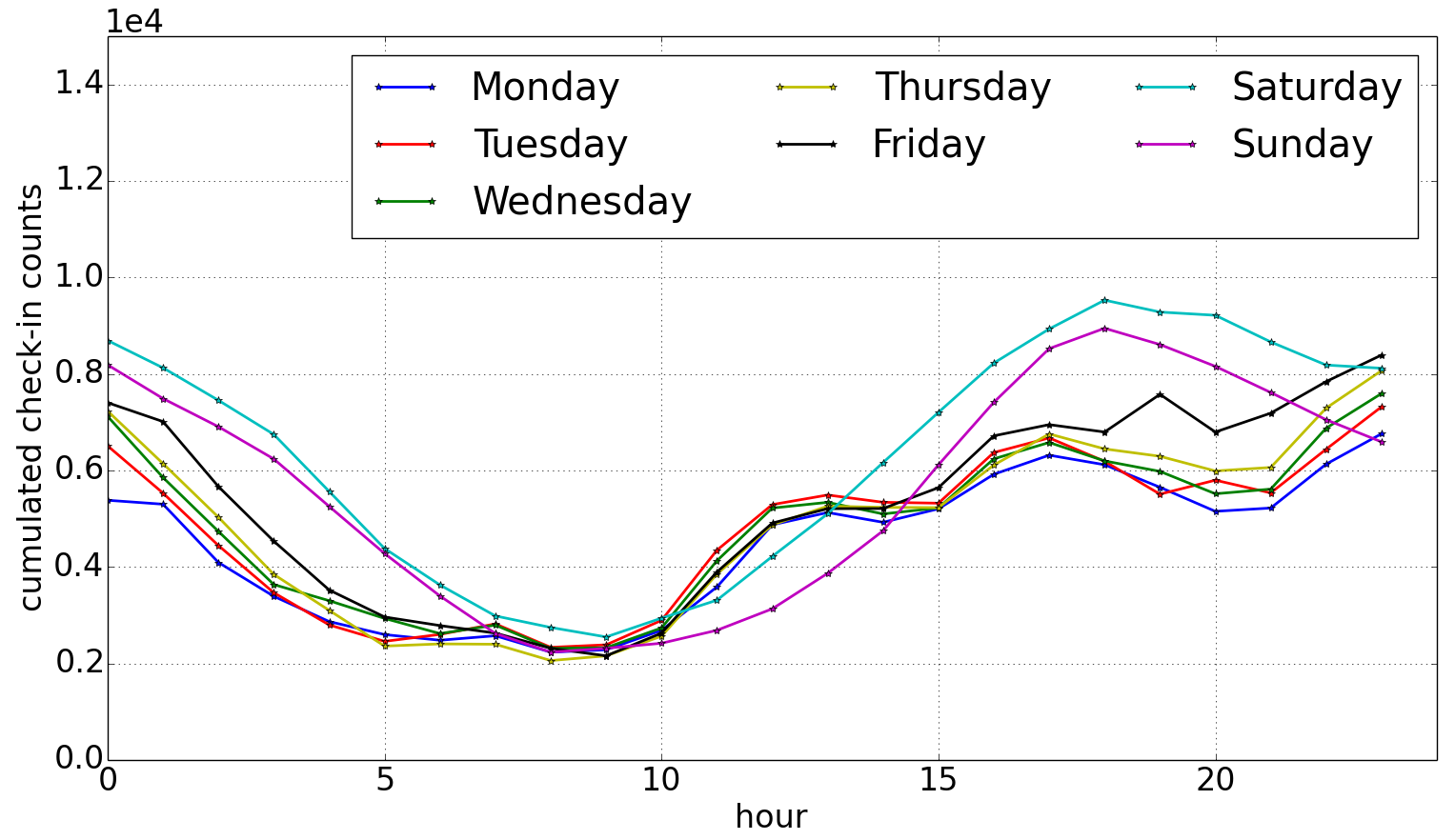}}

\subfigure[Gowalla]{\includegraphics[height=1.8in,width=\columnwidth]{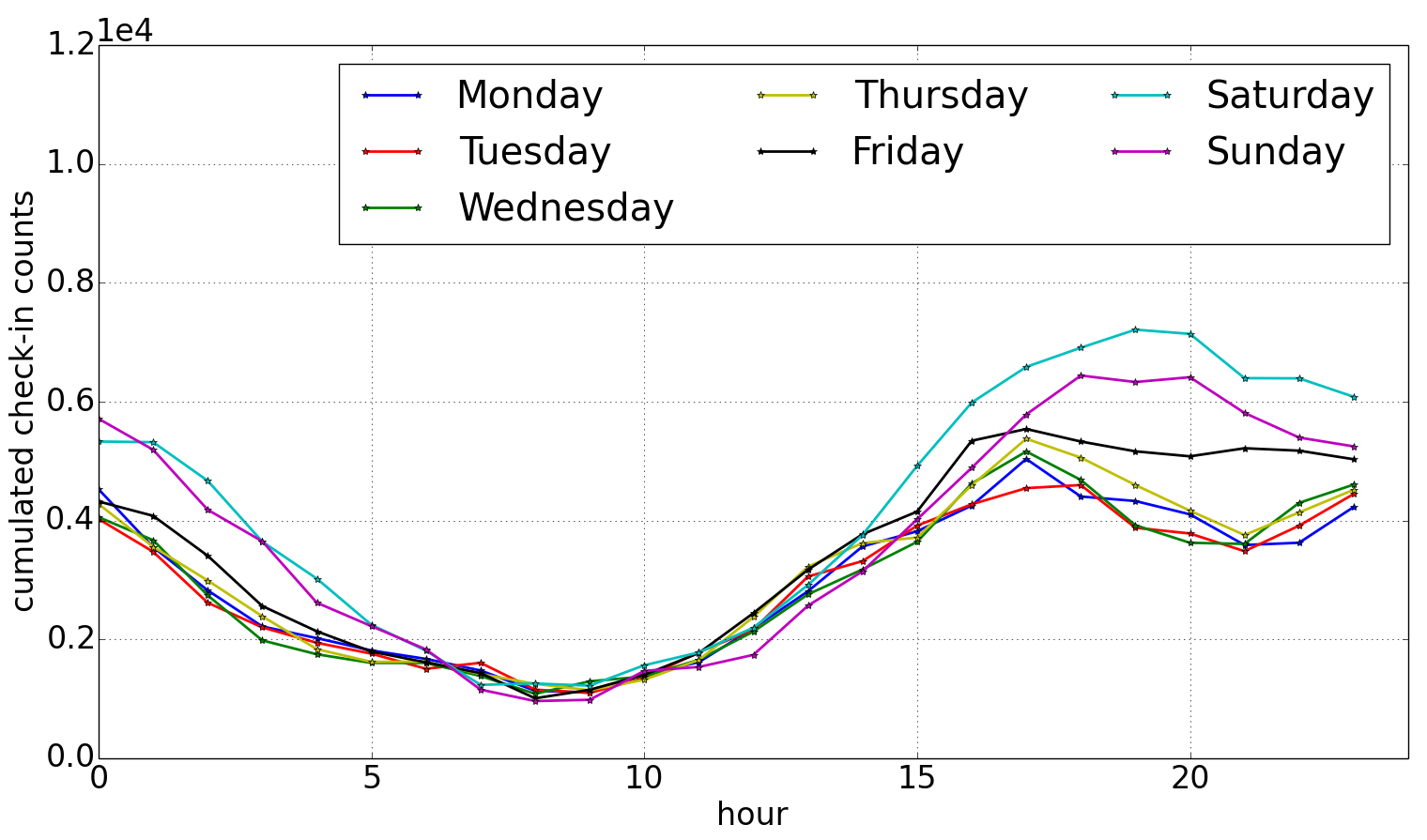}\label{fig:4s_week}}
\caption{Day of week check-in pattern at different hours}
\label{fig:week_day}

\end{figure}

\subsection{Empirical Analysis}
\label{subsec:ea}
We conduct data analysis to answer the following two questions: 1) how POIs in sequences of one day correlate each other? 2) how check-in sequences perform on different days?

We investigate the correlations of POIs in sequences of one day, as shown in Figure~\ref{fig:ccdf}. 
To calculate the correlation between two POIs, we construct the user-POI matrix according to the check-in records. Then, we measure the correlation of a POI pair in terms of the Jaccard similarity of those users who have checked-in at the two POIs. 
In Figure~\ref{subfig:interpoi}, we calculate the average correlation value of POI pairs in sequences for all users, and compare it with average correlation value of 5,000 random POI pairs. We observe that the correlation of POIs in sequences is much higher than random pairs by about 100 times for Foursquare and 50 times for Gowalla, which motivates the sequential modeling. In Figure~\ref{subfig:interpoig}, we compare the correlation of consecutive pairs with nonconsecutive pairs in sequences. Take a sequence of $(l_1, l_2, l_3)$ as an example, $(l_1, l_2)$ and $(l_2, l_3)$ are consecutive pairs, and $(l_1, l_3)$ is a nonconsecutive pair. We also calculate the average value of all sequences for all users to make the comparison. We observe that the nonconsecutive pairs contain comparable correlation to the consecutive pairs. Hence, not only consecutive POIs are highly correlated~\cite{cheng2013you,zhao2016stellar}, all POIs in a sequence are highly correlated with a contextual property. 
Accordingly, it is not satisfactory to only model the consecutive check-ins' transitive probability by Markov chain model or the consecutive check-ins' correlation by tensor factorization. This observation motivates us to model the whole sequence through the \textit{word2vec} framework. 

We explore how the variant temporal characteristics on different days affect the user's check-in behavior. Figure~\ref{fig:week_day} demonstrates
the number of cumulated check-ins for all users at different hours on different days of a week, from Monday to Sunday. From the statistics of cumulated check-ins in Figure~\ref{fig:week_day}, we observe the day of week check-in pattern at different hours: Saturday and Sunday take the similar pattern, while Monday to Friday take an intra similar pattern that is different from the weekends. We may infer that \texttt{weekday} and \texttt{weekend} exert two types of effects on the user's check-in behavior. Therefore, modeling the sequence pattern should contain this temporal feature.

\section{Model}
\label{sec:model}
In this section, we first demonstrate how to capture the sequential pattern through our POI embedding model. Then, we propose the \textit{SEER} model  to learn the POI recommendation system. Next, we propose the temporal POI embedding model and propose the \textit{T-SEER} model to incorporate the temporal influence.  Further, we incorporate the geographical influence into the \textit{T-SEER} model and propose the \textit{GT-SEER} model. Finally, we report how to learn the proposed models. In order to help understand the paper, we list some important notations in the following, shown in Table~\ref{tbl:sym}.

\begin{table}[!t]
\small
\centering
\caption{Symbol notations}
\begin{tabular}{l |l}
\hline
$u$ & user name \\ \hline
$l$ & POI name\\ \hline
$t_s$ & temporal state for a sequence \\ \hline
$k$ & context window size \\ \hline 
$h$ & negative sample size for embedding learning \\ \hline
$m$ & negative sample size for preference learning \\ \hline
$d$ & latent vector dimension \\ \hline
$C$ & the set of  check-ins \\ \hline
$U$ & the set of users    \\ \hline
$L$ & the set of  POIs  \\ \hline
$S_u$ & a sequence for user $u$  \\ \hline
$S$ & the set of sequences  \\  \hline
$D_{S_u}$ & the set of preference relations for $S_u$ \\\hline
$\textbf{T}$ & temporal state feature matrix  \\ \hline
$\textbf{U}$ & user latent feature matrix  \\ \hline
$\textbf{L}$ & user latent feature matrix \\ \hline

\end{tabular}
\label{tbl:sym}
\end{table}


\subsection{POI Embedding}
\label{subsec:tpe}
We propose a POI embedding method to learn the sequential pattern, which captures POIs' contextual information from user check-in sequences.
Our model is based on the \textit{word2vec} framework, i.e., Skip-Gram model~\cite{mikolov2013distributed}. In order to learn the POI representations, we treat each user as a ``document", check-in sequence in a day as a ``sentence", and each POI as a ``word". 
To better describe the model, we present some basic concepts as follows.

\begin{definition} [{check-in}]
 A check-in is a triple $\langle u,l,t \rangle$ that depicts a user $u$ visiting POI $l$ at time $t$. 
\end{definition}

\begin{definition} [{Check-in sequence}]
 A check-in sequence is a set of check-ins of user $u$ in one day, denoted as $S_u =$ $\{ \langle l_1, t_1 \rangle,$ $\dots , $ $\langle l_n, t_n \rangle\} $, where $t_1$ to $t_n$ belong to the same day.  For simplicity, we denote $S_u=\{l_1, \dots, l_n \}.$
\end{definition}

\begin{definition} [{Target POI and context POI}]
 In a sequence $S_u$, the chosen $l_i$ is the target POI and other POIs in $S_u$ are context POIs.  
\end{definition}


%
%

\begin{figure}
[!t] 
\centering
\includegraphics[scale=0.45]{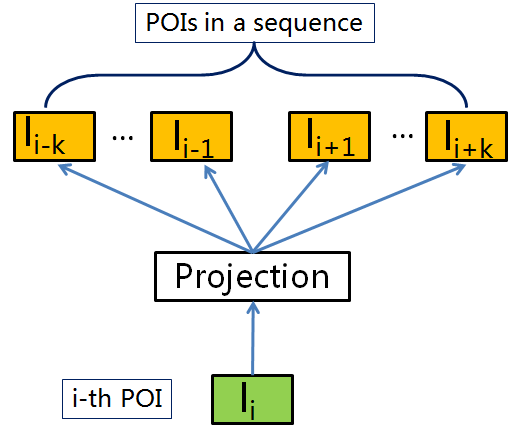}
\caption{POI embedding model}
\label{fig:skip}
\end{figure}
POI embedding model learns the representations from check-in sequences as shown in Figure~\ref{fig:skip}. We treat each POI as a unique continuous vector, and then represents context POIs in a sliding window from $l_{i-k}$ to $l_{i+k}$ given a target POI $l_i$. In other words, the vector of a target POI $l_i$ is used as a feature to predict the context POIs from $l_{i-k}$ to $l_{i+k}$.
Formally, given a POI sequence $S_u=\{l_1, \dots, l_n \},$ the objective function of POI embedding model is to maximize the average log probability,
\begin{equation}
\label{eq:skip}
\small
\mathcal{L}(S_u) = \frac{1}{|S_u|} \sum_{l_i \in S_u} \sum_{-k \leq c \leq k, c \neq 0} \log \Pr(l_{i+c}|l_i),
\end{equation}
where $l_i$ is the target POI, $l_{i+c}$ is the context POI, and $k$ is the context size controlling the sliding window.  Here, we formulate the probability $\Pr(l_{i+c}|l_i)$ using a softmax function. Denote $\textbf{l}'_c, \textbf{l}_i \in R^d$ are the vector representations of output layer context POI $l_{i+c}$  and target POI $l_i$ respectively, $d$ is the vector dimension. Then, the probability $\Pr(l_{i+c}|l_i)$ is formulated as, 
\begin{equation}
\Pr(l_{i+c}|l_i) = \frac{\exp(\textbf{l}'_c \cdot \textbf{l}_i)}{\sum_{l_i \in L} \exp(\textbf{l}'_c \cdot \textbf{l}_i)},
\end{equation}
where $L$ is the POI set and  $(\cdot)$ is the inner product operator.

In order to make the model efficient for learning, Mikolov et al.~\cite{mikolov2013distributed} propose two methods to learn the \textit{word2vec} model, hierarchical softmax and negative sampling.  
In this paper, we employ the negative sampling technique. 
Now we avoid to calculate the softmax function directly. We attempt to maximize the context POI's occurrence and minimize the negative sample's occurrence. 
Then, the objective function could be formulated in a new form easier to optimize. Following~\cite{mikolov2013distributed},  we can define the $\mathcal{L}(S_u)$ through the negative sampling technique,
\begin{equation}
\label{eq:lsu}
\begin{split}
\mathcal{L}(S_u) =  &\frac{1}{|S_u|} \sum_{l_i \in S_u} \sum_{-k \leq c \leq k, c \neq 0} \big (\log \sigma({\textbf{l}}'_c \cdot {\textbf{l}}_i)+\\
&\sum_{h}\mathbb{E}_{k'\sim P_{nc_{i}}}\log  \sigma(-{\textbf{l}}'_{k'} \cdot \textbf{l}_i)\big ),
\end{split}
\end{equation}
where $l_{k'}$ is the sampled negative POI, $h$ is the number of negative samples, $P_{nc_{i}}$ denotes the distribution of POIs not in $S_{u}$, and $\sigma(\cdot)$ is the sigmoid function. $\mathbb{E}_{k'\sim P_{nc_{i}}}(\cdot)$ means to calculate the expectation value for negative sample $l_{k'}$ generated with distribution $P_{nc_{i}}$.    Here we adopt the same strategy in~\cite{mikolov2013distributed} to draw the negative samples, namely using a  unigram distribution raised to the power $\frac{3}{4}$ to construct $P_{nc_{i}}$.

\subsection{\textbf{\textit{SE}}quential  \textbf{\textit{E}}mbedding \textit{\textbf{R}}ank (\textit{SEER}) Model }
We model the user preference in POI recommendation through pairwise ranking.  User check-ins not only contain the sequential pattern, but also imply the user preference. We observe that check-in activity is a kind of implicit feedback, which has been modeled to capture users' preferences on POIs~\cite{Li:2015:RRB:2766462.2767722,lian2014geomf,liu2014exploiting}. To learn this kind of implicit feedback, we leverage the Bayesian personalized ranking criteria~\cite{rendle2009bpr} to model the user check-in activity. Formally, for each check-in $\langle u, l_i \rangle$, we define the pairwise preference order as,
\begin{equation}
l_i >_{u,} l_n,
\end{equation}
where $l_i$ is the checked-in POI and $l_n$ is any other unchecked POI.  The pairwise preference order means user $u$ prefers the checked-in POI $l_i$ than the unchecked POI $l_n$. Supposing that the function $f(\cdot)$ represents user check-in preference score, we model the pairwise preference order by 
\begin{equation}
\label{eq:prob}
\Pr(l_i >_{u} l_n) = \sigma (f(u,l_i) - f(u, l_n)),
\end{equation} 
where $\Pr(l_i >_{u} l_n)$ denotes the probability of user $u$ prefers POI $l_i$ than $l_n$, and $\sigma(\cdot)$ is the sigmoid function. 

Furthermore, we employ the matrix factorization (MF) model~\cite{koren2009matrix} to formulate the preference score function. 
In other words, we are able to use the latent vector inner product to define the score function,
\begin{equation}
\label{eq:score}
f(u,l) = \textbf{u} \cdot \textbf{l},
\end{equation}
where $\textbf{u}, \textbf{l} \in R^d$ are latent vectors of user $u$ and POI $l$, respectively. Thus, the pairwise preference score function can be formulated as,
\begin{equation}
\label{eq:probmf}
\Pr(l_i >_{u} l_n) = \sigma ( \textbf{u}  \cdot \textbf{l}_i- \textbf{u} \cdot \textbf{l}_n ).
\end{equation} 

Suppose $C$ is the set containing all check-ins,  $S$ is the set containing all sequences, $L$ is the set of POIs, and $L_u$ is the checked-in POIs of user $u$. To model the pairwise preference of check-ins in $S_u$, we sample unchecked POIs from $L\setminus L_u$ and construct a pairwise preference set, 
\begin{equation}
D_{S_u} = \{(u, l_i, l_n) | l_i \in S_u, l_n \in L \setminus L_u \}.
\end{equation}

Hence, learning the pairwise preference relations in $S_u$ is equivalent to maximize the log probability of preference pairs in $D_{S_u}$, 
\begin{equation}
\label{eq:ldsu}
\mathcal{L}(D_{S_u}) =  \underset{( u, l_i, l_n) \in D_{S_u}} {\sum}  \log \sigma (\textbf{u} \cdot (\textbf{l}_i - \textbf{l}_n)).
\end{equation} 

Moreover, we propose the \textit{SEER} model to learn the user preference and as well as sequential pattern for POI recommendation together. 
Learning the \textit{SEER} model is equivalent to maximize  $\mathcal{L}(S_u)$ in Eq.~(\ref{eq:lsu}) and $\mathcal{L}(D_{S_u})$ in Eq.~(\ref{eq:ldsu}) together. 
Therefore, the objective function of the \textit{SEER} model can be formulated as,
\begin{equation}
\mathcal{O} = \argmax  \sum_{S_u \in S} \sum_{l_i \in S_u} ( \alpha \mathcal{L}(S_u) + \beta \mathcal{L}(D_{S_u}),
\end{equation}
where $\alpha$ and $\beta$ are the hyperparameters to trade-off the sequential influence and the user preference.

Substituting  $\mathcal{L}(S_u)$ and $\mathcal{L}(D_{S_u})$ with Eq.~(\ref{eq:lsu}) and Eq.~(\ref{eq:ldsu}) respectively, then we can learn the \textit{SEER} model through the following objective function,
\begin{equation}
\small
\label{eq:seerobj}
\begin{split}
 {\argmax}
 \sum_{S_u \in S} \sum_{l_i \in S_u} & \big (\sum_{-k \leq c \leq k, c \neq 0} \alpha \log \sigma({\textbf{l}}'_c \cdot {\textbf{l}}_i)+ \\&
 \sum_{h}\alpha \mathbb{E}_{k'\sim P_{nc_{i}}}\log  \sigma(-{\textbf{l}}'_{k'} \cdot \textbf{l}_i)+  \\&
 \underset{ D_{S_u}} {\sum} \beta \log  (\sigma (\textbf{u} \cdot (\textbf{l}_i - \textbf{l}_n))) \big).
\end{split}
\end{equation}

\subsection{Temporal SEER (T-SEER) model }
\begin{figure}
[!t] 
\centering
\includegraphics[scale=0.45]{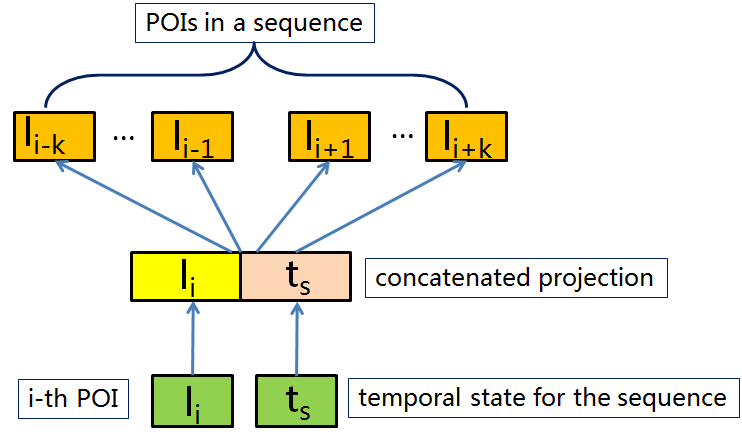}
\caption{Temporal POI embedding model}
\label{fig:tskip}
\end{figure}
 
To model the temporal variance of sequences on different days, we propose the \textit{T-SEER}  model. As shown in Figure~\ref{fig:week_day}, user check-ins demonstrate different patterns on \texttt{weekday} and \texttt{weekend}. Thus, we should model the sequences on \texttt{weekday} and \texttt{weekend} differently. 
The POI embedding model in Figure~\ref{fig:skip} only learns the contextual information of POIs from the check-in sequences, but ignore the variant temporal characteristics among sequences. To this end, we propose the temporal POI embedding model to learn POI representations.

We propose the temporal POI embedding that represents the POI in sequences with specific temporal state. In our case, we want to discriminate  \texttt{weekday} and \texttt{weekend}, hence the temporal state $t_s$ is composed of two options, \texttt{weekday} and \texttt{weekend}.  As shown in Figure~\ref{fig:tskip}, we learn the representations of context POIs from $l_{i-k}$ to $l_{i+k}$ given a target POI $l_i$ and the sequence temporal state $t_s$. Formally, given a sequence $S_u$ and its temporal state $t_s$, our model attempts to learn the temporal POI embeddings through maximizing the following probability,
\begin{equation}
\label{eq:tskip}
\small
\begin{split}
\mathcal{L}(S_u) =& \frac{1}{|S_u|} \sum_{l_i \in S_u} \sum_{-k \leq c \leq k, c \neq 0} \big ( \log \Pr(l_{i+c}|l_i,t_s) \big ).
\end{split}
\end{equation}

Similarly, we formulate the probability $\Pr(l_{i+c}|l_i, t_s)$ using a softmax function. 
For better description, we introduce two symbols, defined as follows: $\hat{\textbf{l}}'_c = \textbf{l}'_c \oplus \textbf{l}'_c$, 
${\textbf{l}}_i^t= \textbf{l}_i \oplus \textbf{t}_s,$ where $\oplus$ is the concatenation operator, and $\textbf{l}'_c$, $\textbf{l}_i$, and $\textbf{t}_s$ are latent vectors of output layer context POI, target POI, and temporal state, respectively.
Thus, we get $\hat{\textbf{l}}'_c \cdot {\textbf{l}}_i^t = \textbf{l}'_c \cdot \textbf{l}_i + \textbf{l}'_c \cdot \textbf{t}_s$.
Therefore, the probability $\Pr(l_{i+c}|l_i, t_s)$ can be formulated as, 
\begin{equation}
\Pr(l_{i+c}|l_i, t_s) = \frac{\exp(\hat{\textbf{l}}'_c \cdot {\textbf{l}}_i^t)}{\sum_{l_i \in L} \exp(\hat{\textbf{l}}'_c \cdot {\textbf{l}}_i^t)}.
\end{equation}

Furthermore, we define the $\mathcal{L}(S_u)$ through the negative sampling technique,
\begin{equation}
\begin{split}
\mathcal{L}(S_u) =  &\frac{1}{|S_u|} \sum_{l_i \in S_u} \sum_{-k \leq c \leq k, c \neq 0} \big (\log \sigma(\hat{\textbf{l}}'_c \cdot {\textbf{l}}_i^t)+\\
&\sum_{h}\mathbb{E}_{k'\sim P_{nc_{i}}}\log  \sigma(-\hat{\textbf{l}}'_{k'} \cdot \textbf{l}_i^t)\big ).
\end{split}
\end{equation}

The key to deducing the temporal pairwise preference ranking is the preference score function. We use ${\textbf{l}}_i^t= \textbf{l}_i \oplus \textbf{t}_s$ to represent the temporal POI latent vector, which is consistent with the temporal POI embedding model. 
In addition, we define $\hat{\textbf{u}} = \textbf{u} \oplus \textbf{u}$, then the score function can be formulated as, 
\begin{equation}
\label{eq:tscore}
f(u,t_s,l_i) = \hat{\textbf{u}} \cdot {\textbf{l}}_i^t.
\end{equation}
Denote the temporal pairwise preference order as $l_i >_{u,t_s} l_n$.  Substituting Eq.~(\ref{eq:tscore}) in Eq.~(\ref{eq:prob}) and eliminating the common term $ \textbf{u} \cdot \textbf{t}_s$,  we get the pairwise preference probability function,
\begin{equation}
\label{eq:tprobmf}
\Pr(l_i >_{u,t_s} l_n) = \sigma (\textbf{u} \cdot (\textbf{l}_i - \textbf{l}_n)).
\end{equation} 

Because Eq.~(\ref{eq:tprobmf}) is equivalent to Eq.~(\ref{eq:probmf}), the objective function $\mathcal{L}(D_{S_u})$ for temporal pairwise preference ranking keeps the same. Therefore, the objective for \textit{T-SEER} model can be formulated as follows, 
\begin{equation}
\small
\label{eq:tseerobj}
\begin{split}
 {\argmax}
 \sum_{S_u \in S} \sum_{l_i \in S_u} &\big (\sum_{-k \leq c \leq k, c \neq 0} \alpha \log \sigma(\hat{\textbf{l}}'_c \cdot {\textbf{l}}_i^t)+ \\&
\sum_{h} \alpha \mathbb{E}_{k'\sim P_{nc_{i}}}\log  \sigma(-\hat{\textbf{l}}'_{k'} \cdot \textbf{l}_i^t)+  \\ &
 \underset{ D_{S_u}} {\sum} \beta \log (\sigma (\textbf{u} \cdot (\textbf{l}_i - \textbf{l}_n))) \big).
\end{split}
\end{equation}

\subsection{\textit{\textbf{Geo-Temporal}} \textbf{\textit{SEER}} (\textit{GT-SEER}) Model}
We propose the \textit{GT-SEER} model by 
incorporating geographical influence. According to Tobler's first law of geography, ``Everything is related to everything else, but near things are more related than distant thing"~\cite{tobler1970computer}. It implies that  POIs adjacent to each other are more correlated, which is verified by observation in prior work~\cite{cheng2012fused,yuan2013time,zhao2013capturing}. Because of the observation that users prefer the POIs nearby the checked-in than POIs far away, we can discriminate the unchecked POIs and reconstruct the pairwise preference set for better preference modeling.

\begin{definition}
[\small{Neighboring POI and non-neighboring POI}]
For each check-in $\langle u,l_i \rangle,$ the neighboring POI is the POI whose distance from $l_i$ is less than or equal to a threshold $s$, while the non-neighboring POI is the POI whose distance is more than $s$. Here the threshold distance $s$ is calculated in kilometer.
\end{definition}

Considering the geographical influence,  each check-in $\langle u,l_i \rangle$  implies two kinds of pairwise preference relations: the user prefers the checked-in POI $l_i$ than the unchecked neighboring POI $l_{ne}$, and prefers the unchecked neighboring POI $l_{ne}$ than the unchecked non-neighboring POI $l_{nn}$.
Denote $d(l_i,l_j)$ as the distance of two POIs $l_i$ and $l_j$, we represent the pairwise preferences for check-in $\langle u, l_i \rangle$ as,
\begin{equation}
l_i >_{u,d(l_i, l_{ne}) \leq s} l_{ne} \hspace{1mm} \vee \hspace{1mm}l_{ne} >_{u, d(l_i, l_{nn}) > s} l_{nn}. 
\end{equation}
Further, we reconstruct the pairwise preference set,
\begin{equation}
\begin{split}
D'_{S_u}  = \{(u, l_i, l_{ne}) \vee (u, l_{ne}, l_{nn})  | (u,l_i) \in C, d(l_i, l_{ne}) \leq s, \\
 d(l_i, l_{nn}) > s, l_{ne},l_{nn} \in L \setminus L_u \}.
 \end{split}
\end{equation}
Finally, we substitute the pairwise preference set in Eq.~(\ref{eq:tseerobj}) to incorporate the geographical influence  and formulate the the objective function of \textit{GT-SEER},
\begin{equation}
\small
\label{eq:geoseerobj}
\begin{split}
\mathcal{O} =  {\argmax}
 \sum_{S_u \in S} \sum_{l_i \in S_u} &\big ( \sum_{-k \leq c \leq k, c \neq 0} \alpha \log \sigma(\hat{\textbf{l}}'_c \cdot {\textbf{l}}_i^t)+ \\&
\sum_{h}\alpha\mathbb{E}_{k'\sim P_{nc_{i}}}\log  \sigma(-\hat{\textbf{l}}'_{k'} \cdot \textbf{l}_i^t)+   \\&
 \underset{ D'_{S_u}} {\sum} \beta \log  (\sigma (\textbf{u} \cdot (\textbf{l}_i - \textbf{l}_n))) \big),
\end{split}
\end{equation}
where we substitute the preference set $D_{S_u}$ with a geographical preference set $D'_{S_u}$, other symbols retain the same as Eq.~(\ref{eq:tseerobj}).

\begin{algorithm}[t!]
\small
\LinesNumbered
\KwIn{$S$}
\KwOut{$\textbf{U}$, $\textbf{L}$, $\textbf{T}$}
    Initialize $\textbf{U}$, $\textbf{L}$, $\textbf{L}'$, and $\textbf{T}$ (uniformly at random)\\
    \For{iterations}
    {
        \For{$S_u \in S$}
        {
            \For{ $\langle u,l_i \rangle \in S_{u}$}
            {
				\For { each context POI $l_c$ }
				{
				  Update parameters according to Eq.~(\ref{eq:embedc}) \\
                \For {$ \{k'\sim P_{nc_{c}}\}$}
                {
              	Update parameters according to Eq.~(\ref{eq:embedn})
				} }           	
            	Uniformly sample $m$ unchecked POIs \\
            	\For {$(u, l_i, l_{ne}) \in D_m$}
            	{
   $\delta = 1-\sigma (\textbf{u} \cdot \textbf{l}_i - \textbf{u} \cdot \textbf{l}_{ne})$   \\
    $\textbf{u} \leftarrow \textbf{u} + \beta \eta \delta (\textbf{l}_i-\textbf{l}_{ne})$ \\
            $\textbf{l}_i \leftarrow \textbf{l}_i +  \beta \eta \delta \textbf{u} $ ;
            $\textbf{l}_{ne} \leftarrow \textbf{l}_{ne} -  \beta \eta \delta \textbf{u} $ 
            	}
            	\For {$(u, l_{ne}, l_{nn}) \in D_m$}
            	{
            	$\delta = (1-\sigma (\textbf{u} \cdot \textbf{l}_{ne} - \textbf{u} \cdot \textbf{l}_{nn}))$ \\
            		$\textbf{u} \leftarrow \textbf{u} + \beta \eta \delta (\textbf{l}_{ne} - \textbf{l}_{nn})$ \\
            		$\textbf{l}_{ne} \leftarrow \textbf{l}_{ne} + \beta \eta \delta \textbf{u} $ ;
            		$\textbf{l}_{nn} \leftarrow \textbf{l}_{nn} - \beta \eta \delta \textbf{u} $ 
            	}
              
            }

        }
    }
    \caption{Model learning of \textit{GT-SEER}.}
    \label{alg:1}
\end{algorithm}
\subsection{Learning}
We use an alternate iterative update procedure and employ stochastic gradient descent to learn the objective function.
The objective function of our model is to optimize two parts together, $\mathcal{O} = \argmax  \sum_{S_u \in S} \sum_{l_i \in S_u} ( \alpha \mathcal{L}(S_u) + \beta \mathcal{L}(D_{S_u})$.  
To learn the model, for each sampled training instance, we separately calculate the derivatives for  $\mathcal{L}(S_u)$ and $\mathcal{L}(D_{S_u})$ and update the corresponding parameters along the ascending gradient direction,
\begin{equation}
\label{eq:gradient}
\Theta^{t+1}=\Theta^{t}+\eta\times\frac{\partial\mathcal{O}(\Theta)}{\partial\Theta},
\end{equation}
where $\Theta$ is the training parameter and $\eta$ is the learning rate.

Specifically, for a check-in $\langle u, l_i\rangle$, we calculate the stochastic gradient decent for $\mathcal{L}(S_u)$. First, we get the updating rule for the context POI $l_c$, 
\begin{equation}
\label{eq:embedc}
\begin{split}
&\textbf{l}_i \leftarrow \textbf{l}_i + \alpha \eta  (1-\sigma(\hat{\textbf{l}}'_c \cdot {\textbf{l}}_i^t))\textbf{l}'_c \\
&\textbf{t}_i \leftarrow \textbf{t}_i + \alpha \eta (1-\sigma(\hat{\textbf{l}}'_c \cdot {\textbf{l}}_i^t))\textbf{l}'_c  \\
&\textbf{l}'_c \leftarrow \textbf{l}'_c + \alpha \eta (1-\sigma(\hat{\textbf{l}}'_c \cdot {\textbf{l}}_i^t)) (\textbf{l}_i+\textbf{t}_i). \\
\end{split}
\end{equation}
Then, we update the negative sample $l'_k$ as follows, 
\begin{equation}
\label{eq:embedn}
\begin{split}
&\textbf{l}_i \leftarrow \textbf{l}_i - \alpha\eta \sigma(\hat{\textbf{l}}'_{k'} \cdot {\textbf{l}}_i^t) \textbf{l}'_{k'} \\
&\textbf{t}_i \leftarrow \textbf{t}_i - \alpha \eta \sigma(\hat{\textbf{l}}'_{k'} \cdot {\textbf{l}}_i^t) \textbf{l}'_{k'}  \\
& \textbf{l}'_{k'} \leftarrow \textbf{l}'_{k'} - \alpha \eta \sigma(\hat{\textbf{l}}'_{k'} \cdot {\textbf{l}}_i^t) (\textbf{l}_i+\textbf{t}_i). \\
\end{split}
\end{equation}
To update $\mathcal{L}(D_{S_u})$, we calculate the stochastic gradient decent for each pair $(u, l_i, l_n)$. Denote $\delta = 1-\sigma (\textbf{u} \cdot \textbf{l}_{i} - \textbf{u} \cdot \textbf{l}_{n})$, we update the parameters as follows,
\begin{equation}
\begin{split}
&\textbf{u} \leftarrow \textbf{u} + \beta \eta \delta (\textbf{l}_{i} - \textbf{l}_{n})\\
&\textbf{l}_{i} \leftarrow \textbf{l}_{i} + \beta \eta \delta \textbf{u}\\
& \textbf{l}_{n} \leftarrow \textbf{l}_{n} - \beta \eta \delta \textbf{u}. \\
\end{split}
\end{equation}

%
 Algorithm~\ref{alg:1} shows the details of learning the \textit{GT-SEER} model. $S$ is the set of all sequences, and $S_u$ is a sequence of user $u$.  $\textbf{U}$, $\textbf{L}$, and $\textbf{T}$ are feature matrices of user, POI, and temporal state. $\textbf{L}'$, an assistant learning parameter,  is the output layer POI matrix in Skip-Gram model. We use the standard way~\cite{mikolov2013distributed} to learn the POI representations in the sequences, as shown from line 5 to line 10 in Algorithm~\ref{alg:1}. Next, we exploit the Bootstrap sampling to generate $m$ unchecked POIs and then classify the unchecked POIs as neighboring POIs and non-neighboring POIs according to their distances from the checked-in POI $l_i$. Then, we establish the pairwise preference set $D_m$ for each check-in $\langle u, l_i \rangle$. Here 
$D_m  = \{(u, l_i, l_{ne}) \vee (u, l_{ne}, l_{nn})  |  d(l_i, l_{ne}) \leq s,
 d(l_i, l_{nn}) > s, l_{ne},l_{nn} \in L \setminus L_u \}.$
  Then we learn the parameters for each instance in $D_m$, shown from line 12 to line 21 in Algorithm~\ref{alg:1}. 
Here, we show the detailed updating rules for \textit{GT-SEER} model. The \textit{SEER} model and \textit{T-SEER} model are special cases of the \textit{GT-SEER} model, so we can use similar means to learn them.

After learning the \textit{GT-SEER} model, we get the latent feature representations of users, POIs, and temporal states. Then we can estimate the check-in possibility of user $u$ over a candidate POI $l$ at temporal state $t_s$ according to the preference score function. For \textit{SEER} model, we use the Eq.~(\ref{eq:score}) to estimate the check-in possibility. For \textit{T-SEER} model and \textit{GT-SEER} model, we use the  Eq.~(\ref{eq:tscore}) for score estimation. Finally, we rank the candidate POIs and select the top $N$ POIs with the highest estimated possibility values for each user. 

\textbf{Scalability.} After using some sampling techniques, the complexity of our model is linear in $O(|C|),$ where $C$ is the set of all check-ins. Hence, this proposed algorithm is scalable. Specifically, the parameter update in Eq.~(\ref{eq:embedc}) and  Eq.~(\ref{eq:embedn}) is in $O(d)$, where $d$ is the latent vector dimension. Hence for each context, the update procedure is in $O(d) + O(h \cdot d)$, where $h$ is the number of negative samples. Because the context sliding window size is $k$, POI embedding learning for each check-in $\langle u, l_i \rangle$ from line 5 to 10 is in $O(k \cdot h \cdot d)$.
For the pairwise preference learning from line 11 to 21, we sample $m$ unchecked POIs, which can generate maximum $O(m^2)$ pairwise preference tuples. For each tuple, the update procedure is in $O(d)$. As a result, the parameter update from line 11 to 21 is in $O(m^2 \cdot d)$. 
Because we employ embedding learning and pairwise preference learning for each check-in, the complexity of our model is $O\big((k \cdot h  + m^2 ) \cdot d\cdot |C| \big)$, where $C$ is the set of all check-ins. For $k$, $h$, $m$, and $d$ are fixed hyperparameters, the proposed model can be treated as linear in $O(|C|)$.
Furthermore, in order to make our model more efficient, we turn to the asynchronous version of stochastic gradient descent (ASGD)~\cite{recht2011hogwild}.
As the check-in frequency distribution of POIs in LBSNs follows a power law~\cite{ye2011exploiting}, this results in a long tail of infrequent POIs, which guarantees to employ the ASGD to parallel the parameter updates. 

\section{Experimental Evaluation}
\label{sec:ee}
We conduct experiments to seek the answers of the following questions: 1) how the proposed models perform comparing with other state-of-the-art recommendation methods? 2) how each component (i.e., sequential modeling, temporal effect, and geographical influence) affects the model performance? 3) how the parameters affect the model performance?

\subsection{Experimental Setting}
Two real-world datasets are used in the experiment: one is from Foursquare provided in~\cite{gao2012gscorr} and the other is from Gowalla in~\cite{zhao2013capturing}. Table~\ref{tbl:fstat} demonstrates the statistical information of the datasets. In order to make our model satisfactory to the scenario of recommending for future check-ins, we choose the first 80\% of each user's check-in as training data, the remaining 20\% for test data, following~\cite{cheng2013you,zhang2014lore}. 
\subsection{Performance Metrics}
In this work, we compare the model performance through \textit{precision} and \textit{recall}, which are generally used to evaluate a POI recommendation system~\cite{gao2013exploring,Li:2015:RRB:2766462.2767722}. 
To evaluate a top-$N$ recommendation system, we denote the \textit{precision} and \textit{recall} as P@$N$ and R@$N$, respectively. 
Supposing $L_{visited}$ denotes the set of correspondingly visited POIs in the test data, and  $L_{N, rec}$ denotes the set of recommended POIs, the definitions of P@$N$ and R@$N$ are formulated as follows,
\begin{equation}
\label{eq:prec}
P@N = \frac{1}{|{U}|} \sum_{u\in U} \frac{|L_{visited} \cap L_{N, rec}|}{ N},
\end{equation}
\begin{equation}
\label{eq:recall}
R@N = \frac{1}{|{U}|} \sum_{u \in U} \frac{|L_{visited} \cap L_{N, rec}|}{|L_{visited}|}.
\end{equation}

\subsection{Model Comparison}
In this paper, we propose three models: \textbf{\textit{SEER}} , \textbf{\textit{T-SEER}}, and \textbf{\textit{GT-SEER}}, with features shown in Table~\ref{tbl:modelfeature}. The \textit{SEER} model captures the sequential influence and user preference, showing advantages of our embedding methods. Temporal influence and geographical influence are important for POI recommendation, and are usually modeled to improve the POI recommendation performance. Hence, we incorporate the temporal and geographical influence into the \textit{SEER} model to establish the  {\textit{T-SEER}} and    \textit{GT-SEER} model. 

\begin{table}[t!]
\center
\caption{Model feature demonstration}
\begin{tabular}{c|c|c|c} \hline
& \textbf{\textit{SEER}} & \textbf{\textit{T-SEER}} &\textbf{\textit{GT-SEER}} \\ \hline
 Pairwise Preference & $\surd$ & $\surd$ & $\surd$ \\ \hline
 Sequential Modeling  & $\surd$ & $\surd$&  $\surd$\\ \hline
Temporal Effect & &$\surd$ & $\surd$   \\ \hline
Geographical Influence  &  &  & $\surd$  \\ \hline
\end{tabular}
\label{tbl:modelfeature}
\end{table}

We compare our proposed models with state-of-the-art collaborative filtering models for implicit feedback and  POI recommendation methods. 
\begin{compactitem}
\item \textit{\textbf{BPRMF}}~\cite{rendle2009bpr}: \textit{ \textbf{B}}ayesian \textit{\textbf{P}}ersonalized \textit{\textbf{R}}anking \textit{\textbf{M}}atrix \textit{\textbf{F}}ac-torization (\textit{BPRMF}) is a popular pairwise ranking method that models the implicit feedback data to recommend top-$k$ items.

\item \textit{\textbf{WRMF}}~\cite{hu2008collaborative,pan2008one}: \textit{ \textbf{W}}eighted \textit{\textbf{R}}egularized \textit{\textbf{M}}atrix \textit{\textbf{F}}actorization (\textit{WRMF}) model is designed  for implicit feedback ranking problem. We set the weight mapping function of user $u_i$ at POI $l_j$ as $w_{i,j} = (1+ 10 \cdot C_{i,j})^{0.5}$, where $C_{i,j}$ is the check-in counts, following the setting in~\cite{liu2014exploiting}.

\item \textit{\textbf{LRT}}~\cite{gao2013exploring}:  \textit{\textbf{L}}ocation \textit{\textbf{R}}ecommendation framework with \textit{\textbf{T}}emporal effects model (\textit{LRT})  is a state-of-the-art POI recommendation method, which captures the temporal effect in POI recommendation.
\item \textit{\textbf{LORE}}~\cite{zhang2014lore}:  \textit{LORE} is state-of-the-art model that exploits the sequential influence for location recommendation. Compared with other work~\cite{cheng2013you,ye2013s}, \textit{LORE} employs the whole sequence's contribution, not only the successive check-ins sequential influence.
\item \textit{\textbf{Rank-GeoFM}}~\cite{Li:2015:RRB:2766462.2767722}:  \textit{Rank-GeoFM }is a ranking based geographical factorization method, which incorporates the geographical and temporal influence in a latent ranking model. 
\end{compactitem}

\subsection{Experimental Results}
In the following, we demonstrate the experimental results on P@$N$ and R@$N$.
Since the models' performances are consistent for different values of $N$, e.g., 1, 5, 10, and 20, we show representative results at 5 and 10 following~\cite{gao2013exploring,gao2015content}. 
For the MF-based baseline methods (i.e., BPRMF, WRMF, LRT, and Rank-GeoFM) and our proposed models, the recommendation performance and the computation cost consistently increase with the latent vector dimension. To be fair, we set the same dimension for all these models.  
In our experiments, we set the latent vector dimension as 50 for the trade-off of computation cost and model performance.
\begin{figure}[t!]%
\centering
\subfigure[Precision-Foursquare]{
\includegraphics[width=1.72in , height = 1.4in]{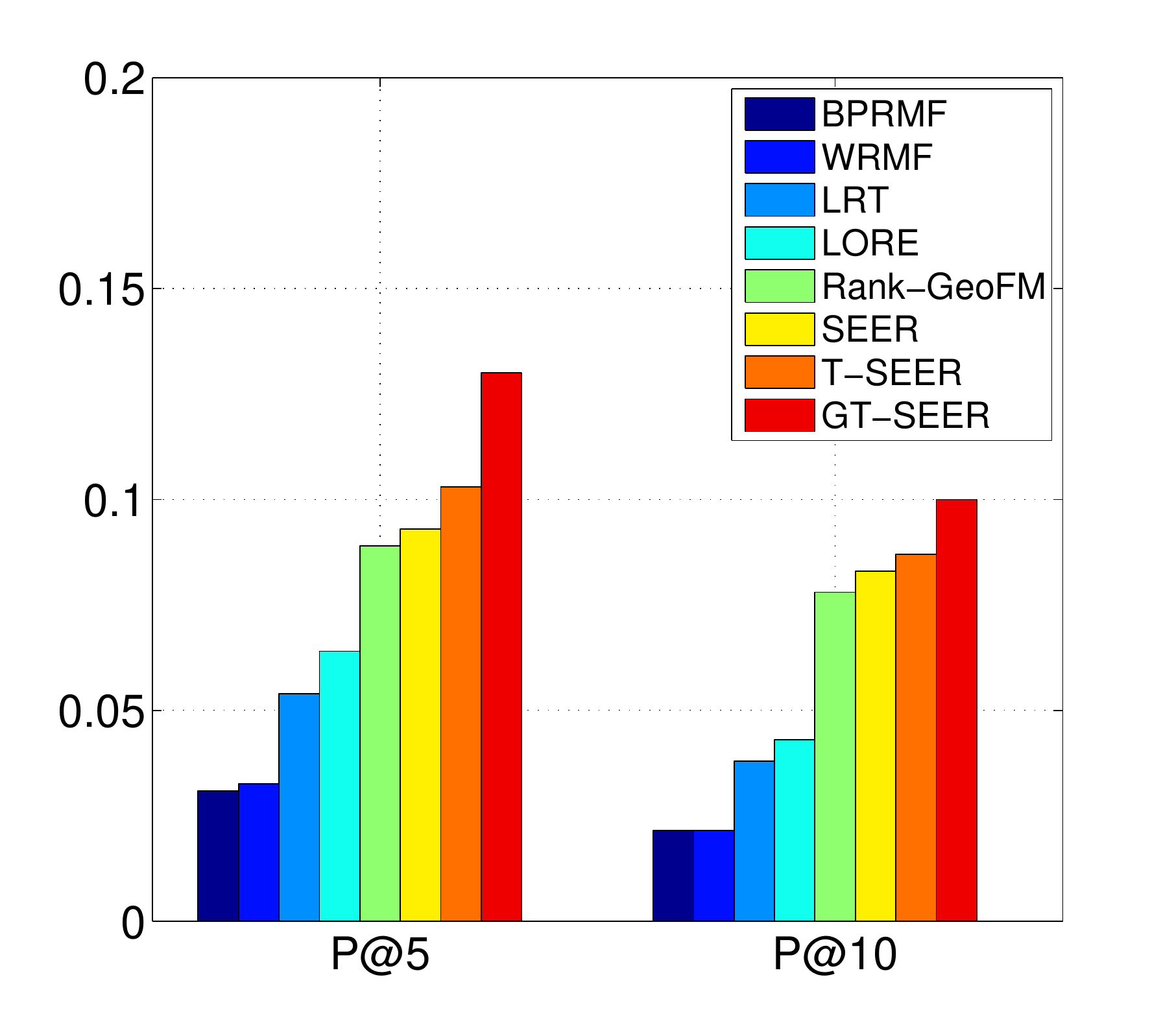}%
\label{subfig:pf}%
}
\subfigure[Recall-Foursquare]{
\includegraphics[width=1.72in , height = 1.4in]{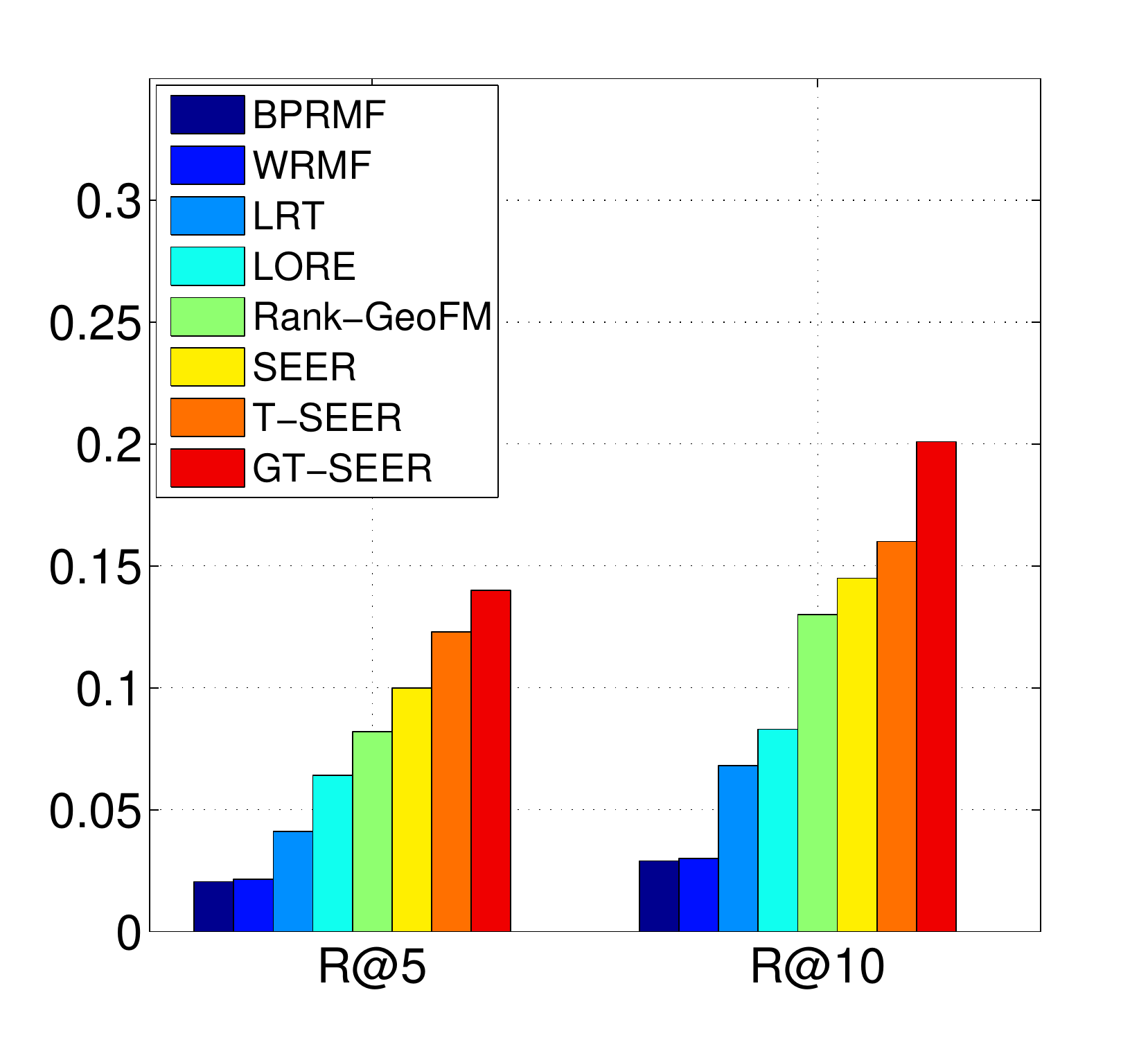}%
\label{subfig:rf}%
}
\subfigure[Precision-Gowalla]{
\includegraphics[width=1.72in , height = 1.4in]{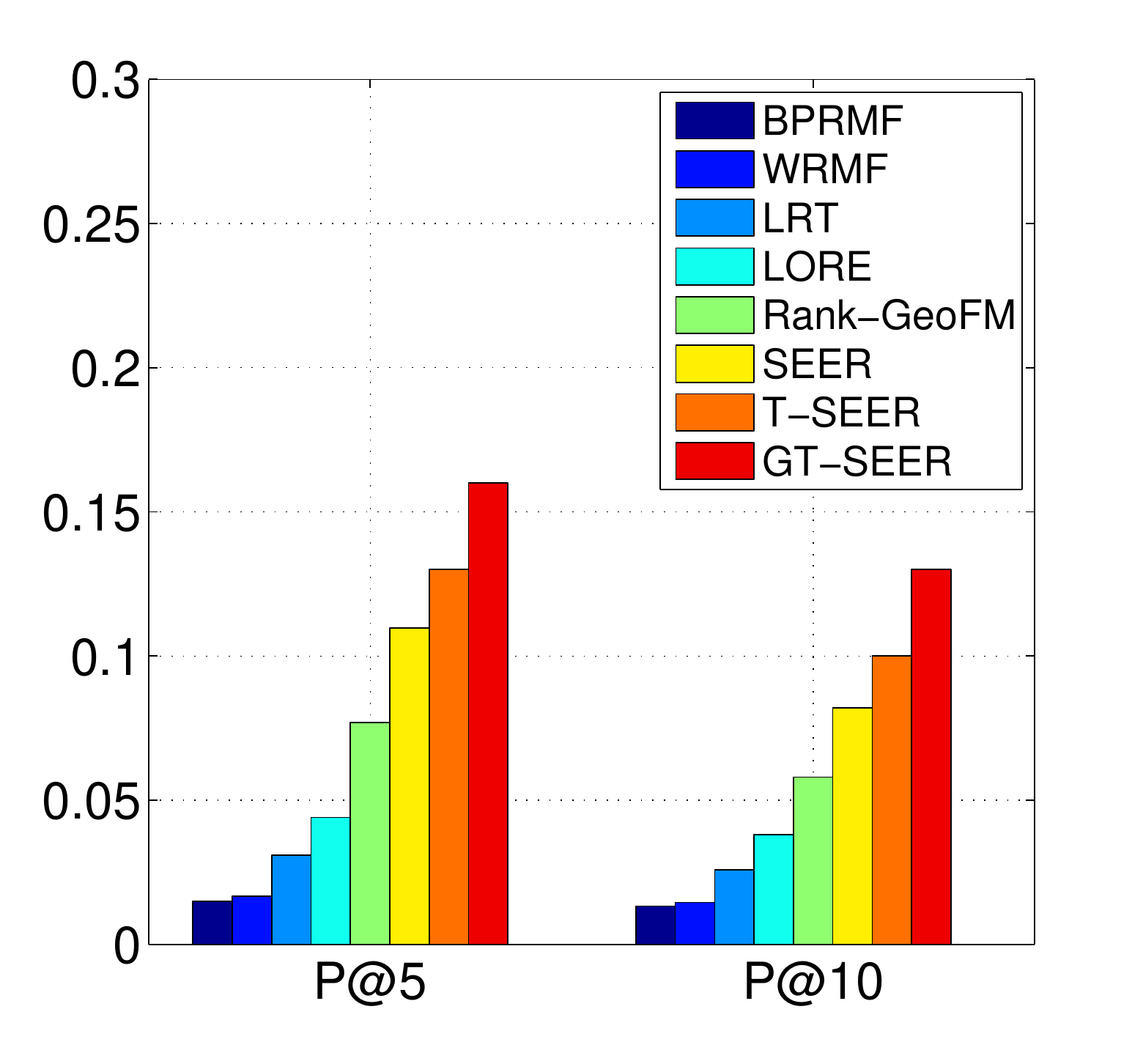}%
\label{subfig:pg}%
}
\subfigure[Recall-Gowalla]{
\includegraphics[width=1.72in , height = 1.4in]{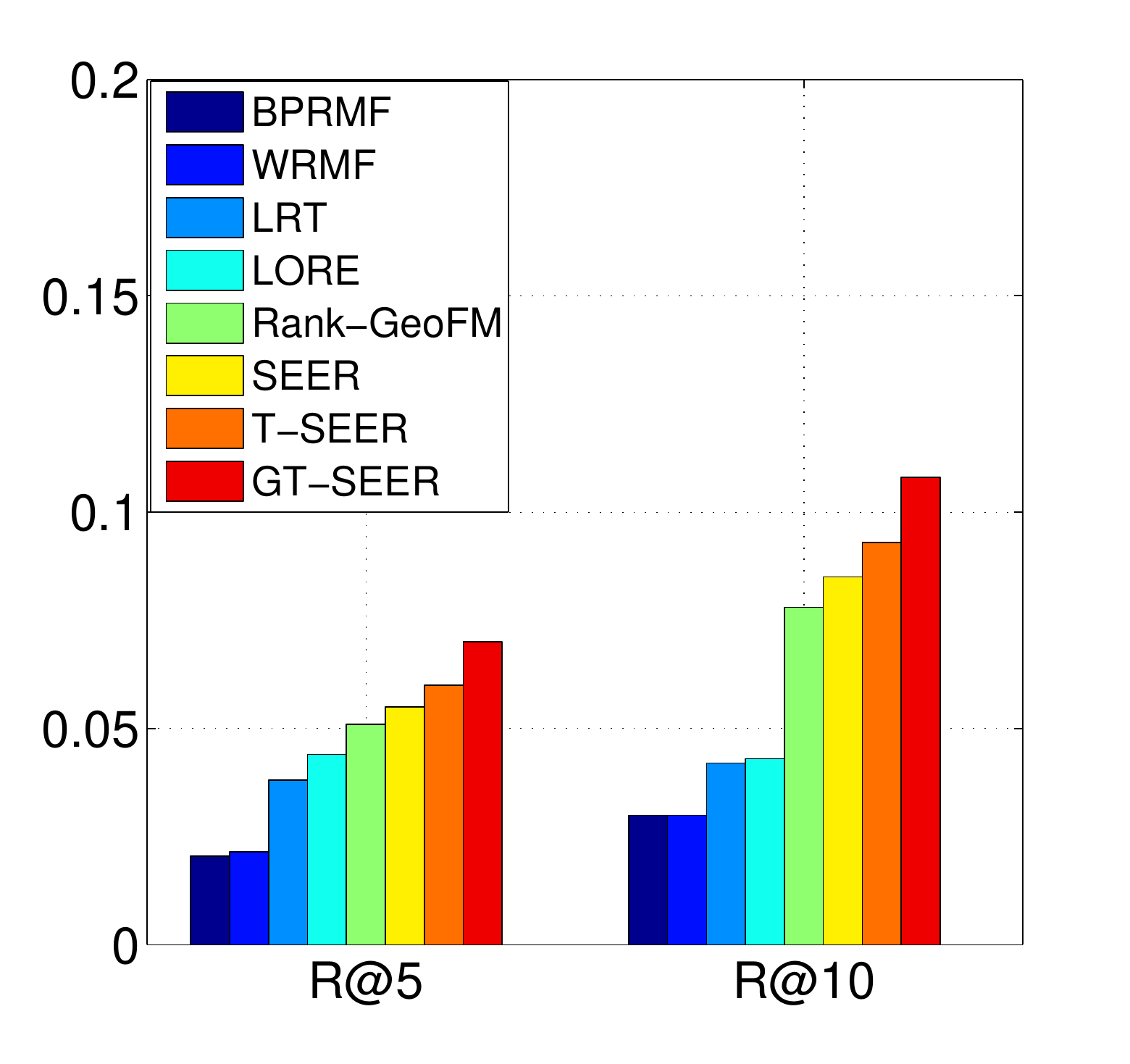}%
\label{subfig:rg}%
}
\caption{Model comparison}
\label{fig:gw_result}
\end{figure}

\subsubsection{Performance Comparison}
From the experimental results, we discover that our proposed models achieve better performance than the baselines, as shown in Figure~\ref{fig:gw_result}. \textit{Rank-GeoFM} is the best baseline competitor. 
Since \textit{Rank-GeoFM} has incorporated the geographical influence and temporal influence, in order to make the comparison fair, we compare \textit{GT-SEER} with \textit{Rank-GeoFM}. 
Experimental results show that \textit{GT-SEER} attains improvements over \textit{Rank-GeoFM} at least 28\% on both datasets for all metrics. This verifies the effectiveness of our sequential modeling and as well as the validity of means for incorporating temporal influence and geographical influence.    
In addition, we observe that models perform better on Gowalla than Foursquare for \textit{precision}, but worse for \textit{recall}. 
The reason lies in that  each user's test data size in Gowalla is bigger than Foursquare. As shown in Table~\ref{tbl:fstat}, the average check-ins for each user in Gowalla is about two times of Foursquare. According to the metrics in Eq.~(\ref{eq:prec}) and Eq.~(\ref{eq:recall}), the result is reasonable.


\subsubsection{Comparison Discussion}
Through the model comparison in Figure~\ref{fig:gw_result},  we verify the strategy of our proposed models, and show the contribution of each component, including sequential modeling, temporal effect, and geographical influence. 

\textbf{\textit{SEER} vs. \textit{BPRMF}. } \textit{BPRMF} is a special case of \textit{SEER} model, when not considering the sequential influence.
The \textit{SEER} model gains more than 150\% improvement on both datasets for all metrics over \textit{BPRMF}. This implies that the sequential influence is important for POI recommendation and our embedding method performs excellently for sequential modeling.

\textbf{\textit{SEER} vs. \textit{LORE}. } The \textit{SEER} model outperforms \textit{LORE} more than 50\%, which indicates our model better captures the sequential pattern. Compared with \textit{LORE}, the \textit{SEER} model takes two advantages: the \textit{word2vec} framework captures the POI contextual information in sequences, and  the sequential correlations and the pairwise preference are jointly learned rather than separately modeled. 

\textbf{\textit{T-SEER} vs. \textit{SEER}. }
The \textit{T-SEER} model captures not only POIs' correlation in a sequence but also the temporal variance in sequences. We observe that \textit{T-SEER} model improves  \textit{SEER} at least about 10\% on both datasets for all metrics. 

\textbf{\textit{GT-SEER} vs. \textit{T-SEER}. }  \textit{GT-SEER} improves the \textit{T-SEER} model at least about 15\% on both datasets for all metrics.  It means our strategy of incorporating geographical influence by discriminating the unchecked POIs is valid.

\begin{figure}[t!]%
\vspace{-2mm}
\centering
\subfigure[P@5 on \textit{SEER}]{
\includegraphics[width=1.7in , height = 1.1in]{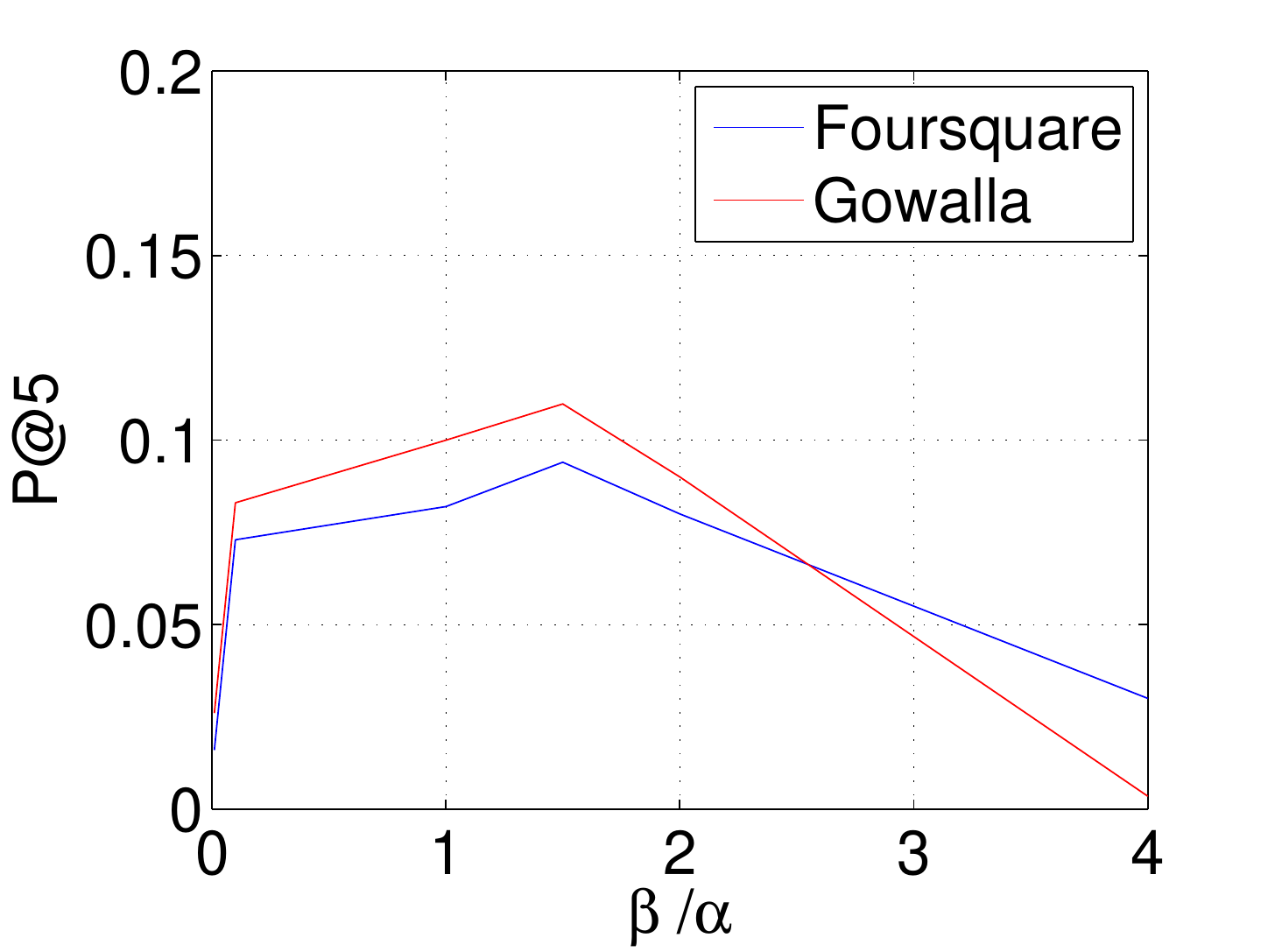}%
\label{subfig:alpha1}%
}\hfill%
\subfigure[R@5 on \textit{SEER}]{
\includegraphics[width=1.7in , height = 1.1in]{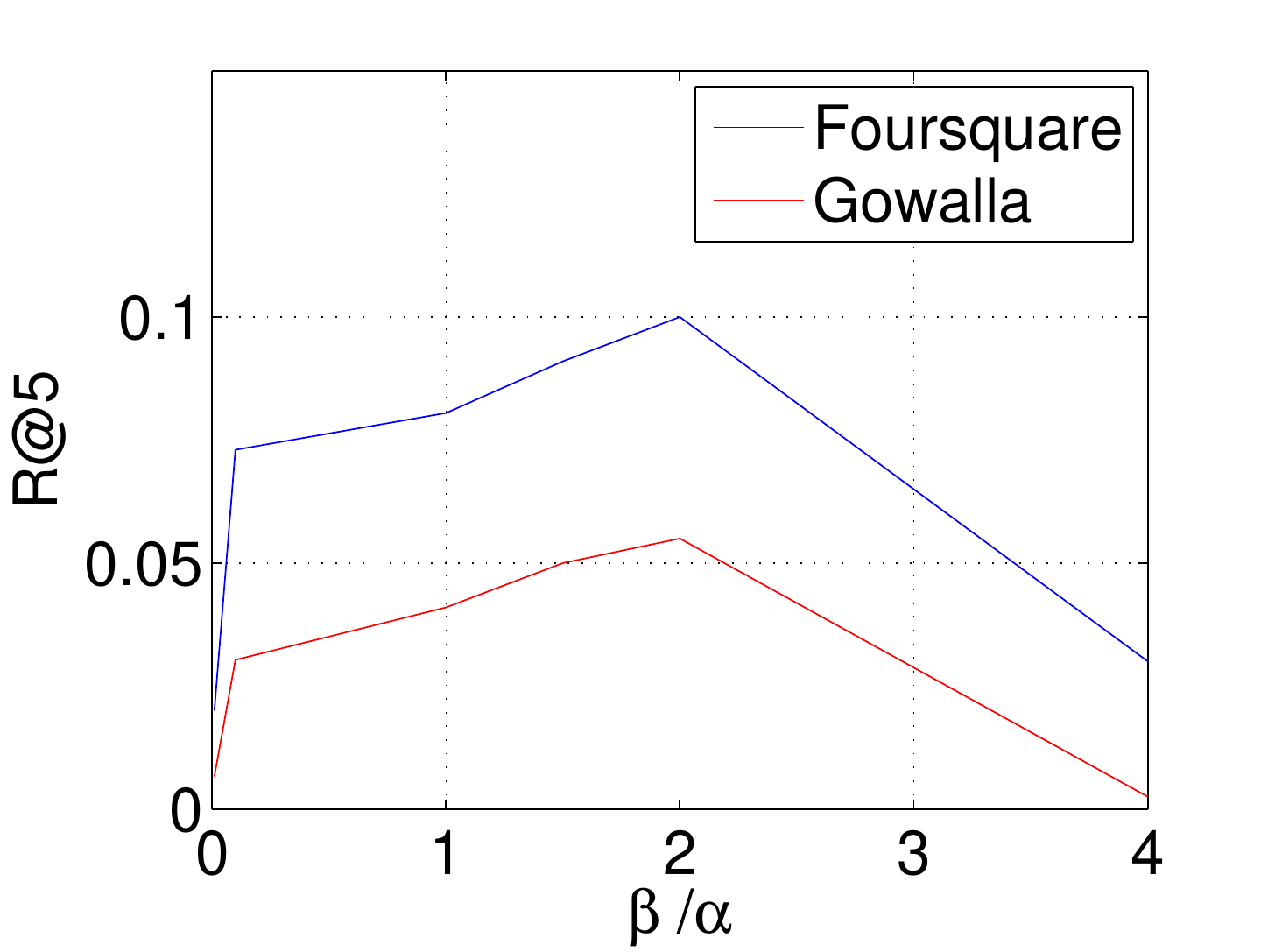}%
\label{subfig:alpha4}%
}\hfill%
\subfigure[P@5 on \textit{T-SEER}]{
\includegraphics[width=1.7in , height = 1.1in]{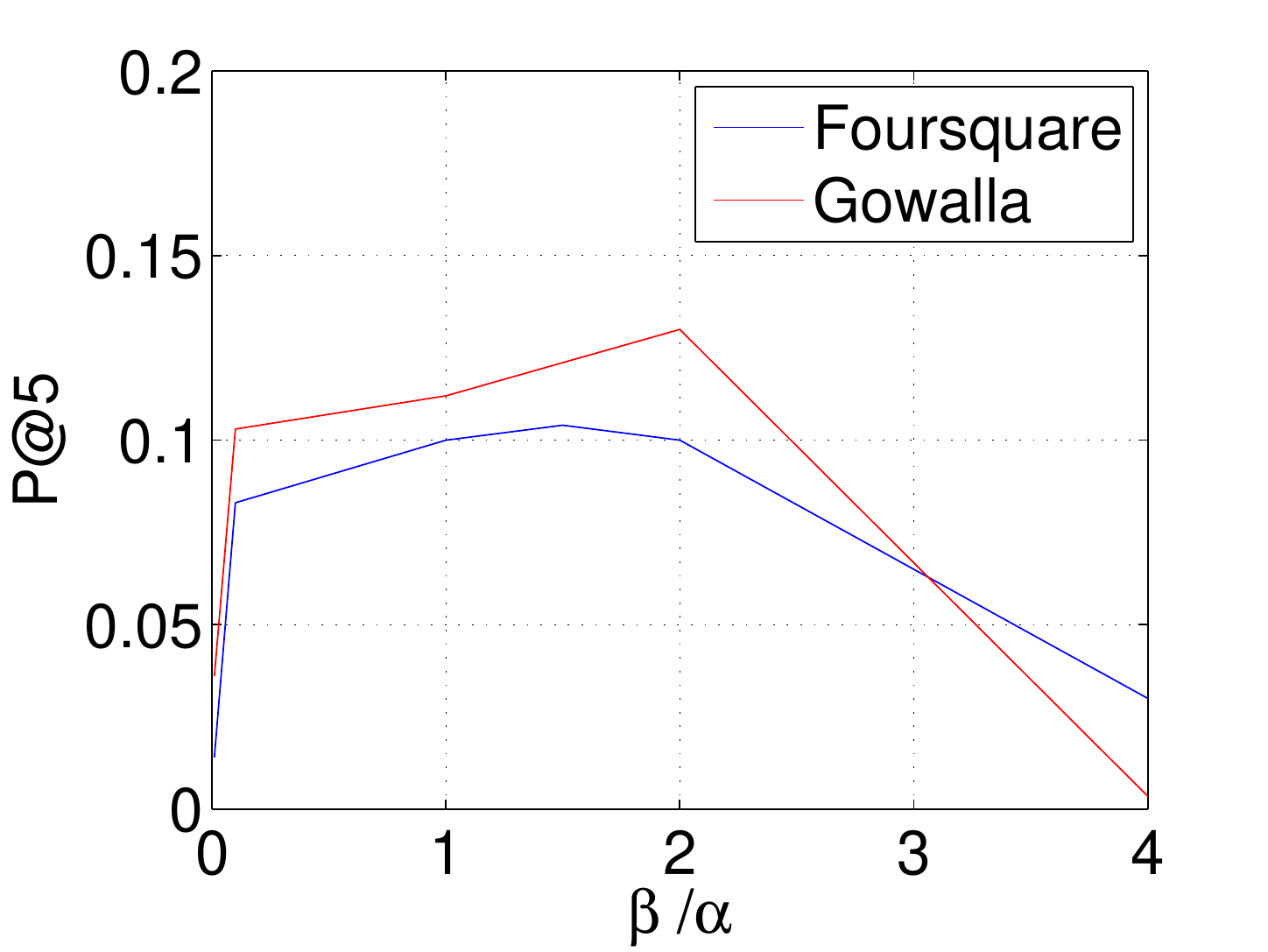}%
\label{subfig:alpha2}%
}\hfill%
\subfigure[R@5 on \textit{T-SEER}]{
\includegraphics[width=1.7in , height = 1.1in]{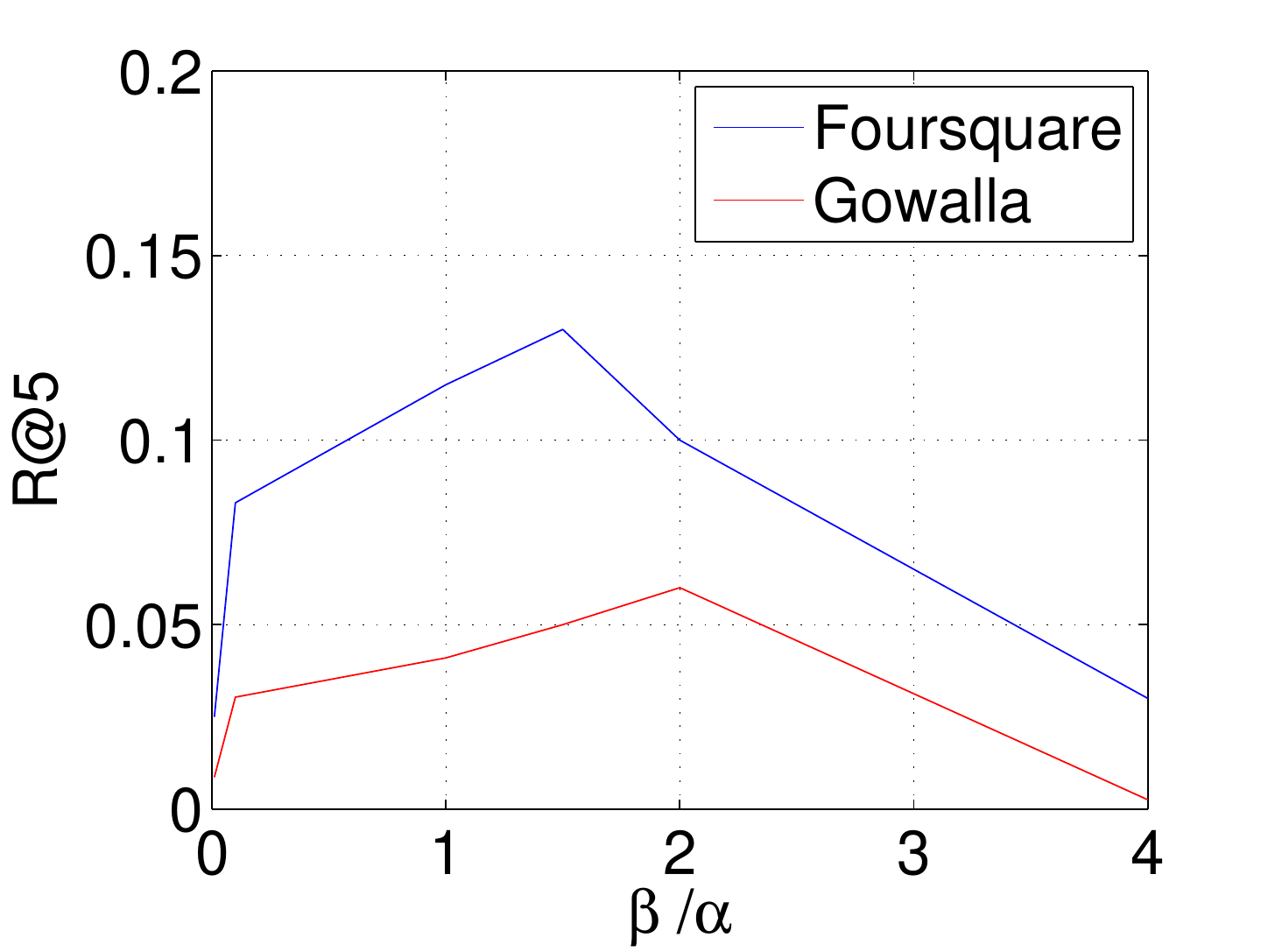}%
\label{subfig:alpha5}%
}\hfill%
\subfigure[P@5 on \textit{GT-SEER}]{
\includegraphics[width=1.7in , height = 1.1in]{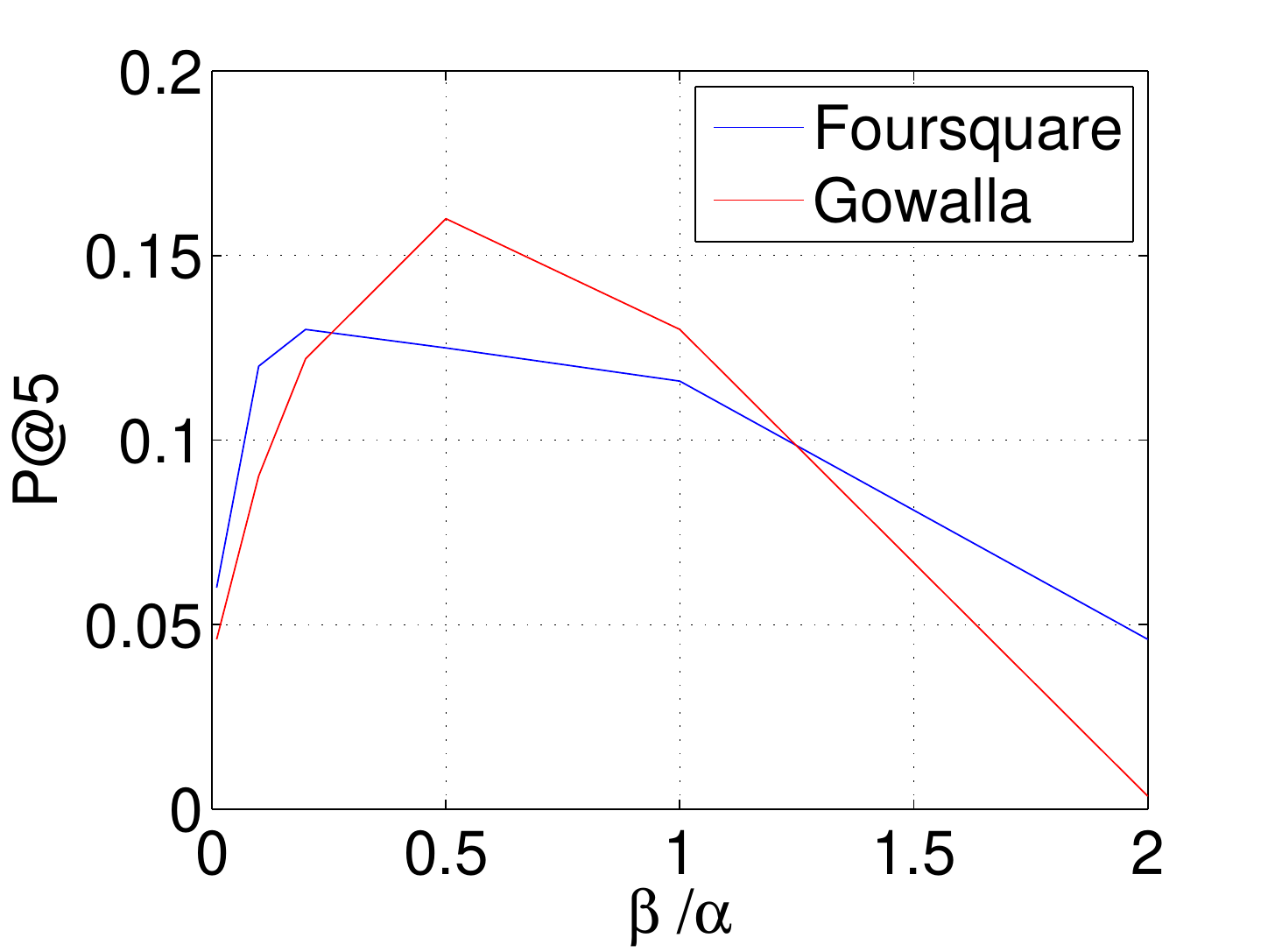}%
\label{subfig:alpha3}%
}
\subfigure[R@5 on \textit{GT-SEER}]{
\includegraphics[width=1.7in , height = 1.1in]{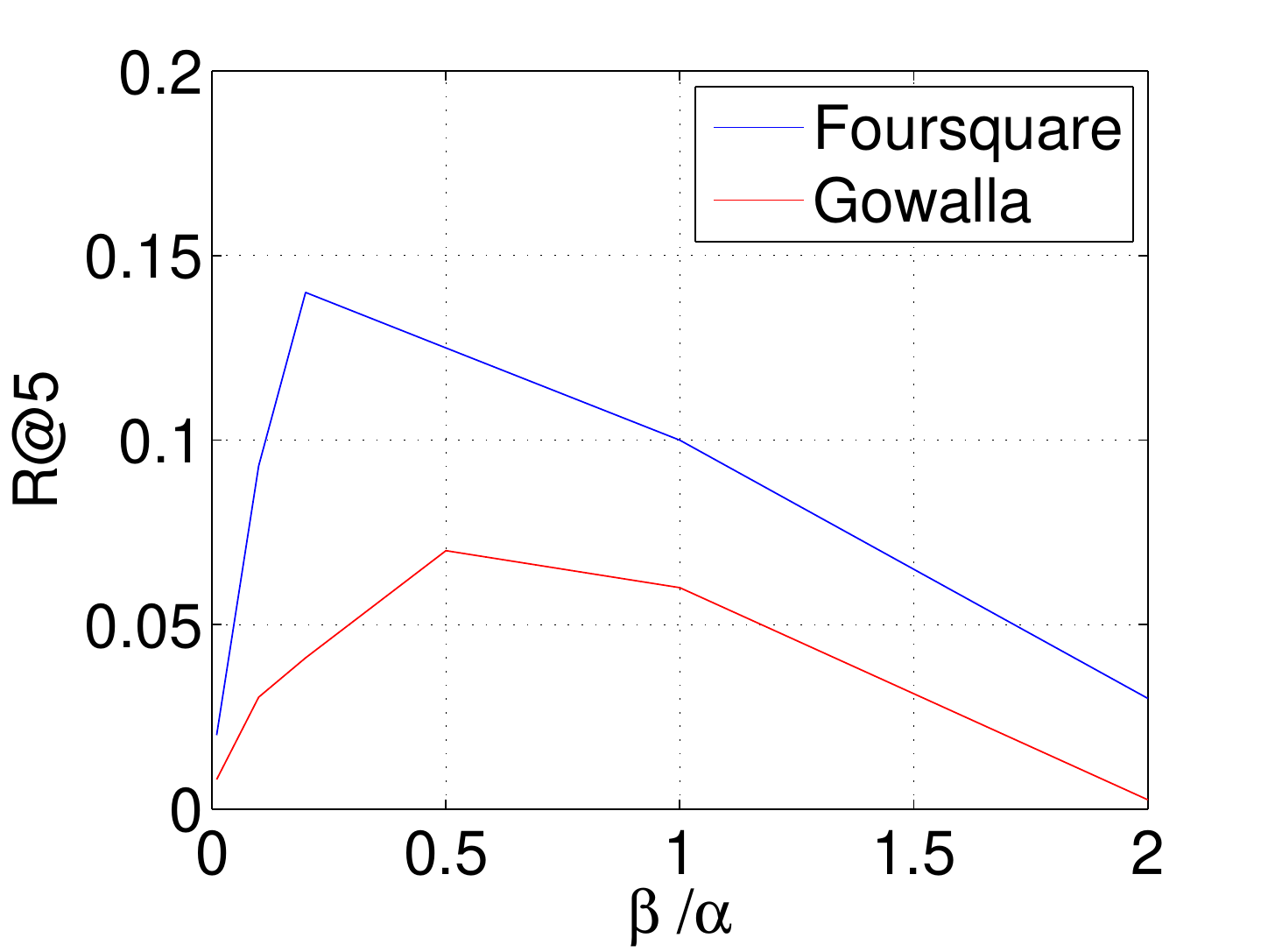}%
\label{subfig:alpha6}%
}
\caption{Parameter effect on $\alpha$ and $\beta$}
\label{fig:alphaparameter}
\vspace{-2mm}
\end{figure}

\subsubsection{Parameter Effect}
In this section, we show how the three important hyperparameters, $\alpha$, $\beta$, and $s$ affect the model performance.  $\alpha$ and $\beta$ balance the sequential influence and the user preference. 
 $s$ shows the sensitivity of our geographical model.

We tune $\alpha$ and $\beta$ to see how to trade-off the sequential influence and user preference, shown in Figure~\ref{fig:alphaparameter} (we only show P@$5$ and R@$5$ for space limit). Both $\alpha$ and $\beta$ appear together with the learning rate $\eta$ in the parameter update procedures. It is not necessary to separately tune the three parameters. We are able to absorb the learning rate $\eta$ into $\alpha$ and $\beta$. In other words, we set 
$\alpha \leftarrow \alpha \cdot \eta,   \beta \leftarrow \beta \cdot \eta$. 
We avoid to tune the learning rate $\eta$, and turn to control the update step size through tuning  $\alpha$ and $\beta$. Hence $\alpha$ and $\beta$ should be small enough to guarantee convergence. We set $\alpha=0.05$, and change $\beta$ to see how the model performance varies with $\frac{\beta}{\alpha}.$ \textit{SEER} and \textit{T-SEER} attain the best performance if $\frac{\beta}{\alpha} \in [1,2]$, while \textit{GT-SEER} attains the best performance if $\frac{\beta}{\alpha} \in [0.25,0.5].$ For \textit{GT-SEER}, more preference pairs are leveraged to train the model such that we need smaller $\beta$ to rebalance the sequential influence and user preference.

\begin{figure}[t!]%
\centering
\subfigure[P@5]{
\includegraphics[width=1.7in , height = 1.1in]{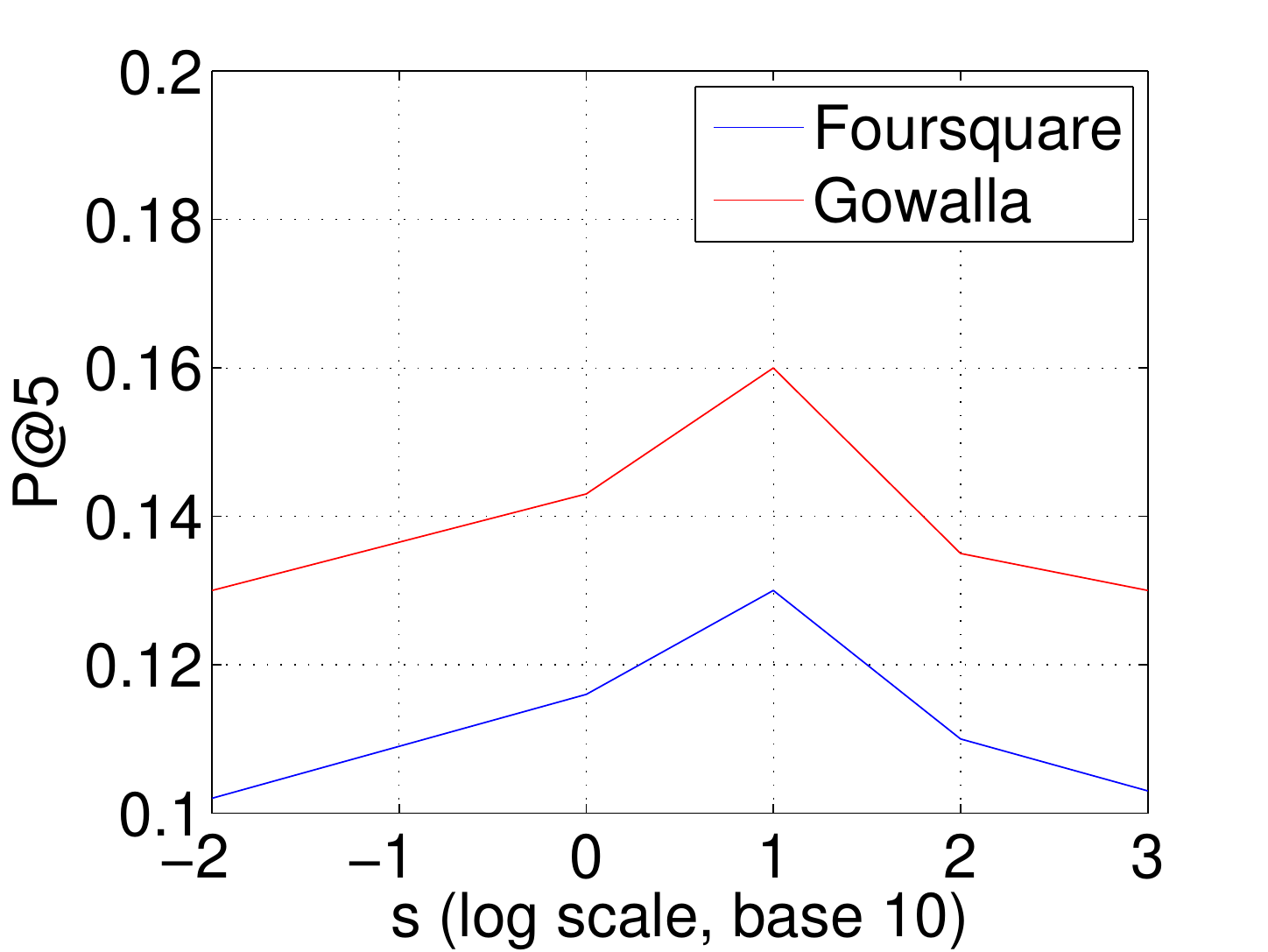}%
\label{subfig:s1}%
}
\subfigure[R@5]{
\includegraphics[width=1.7in , height = 1.1in]{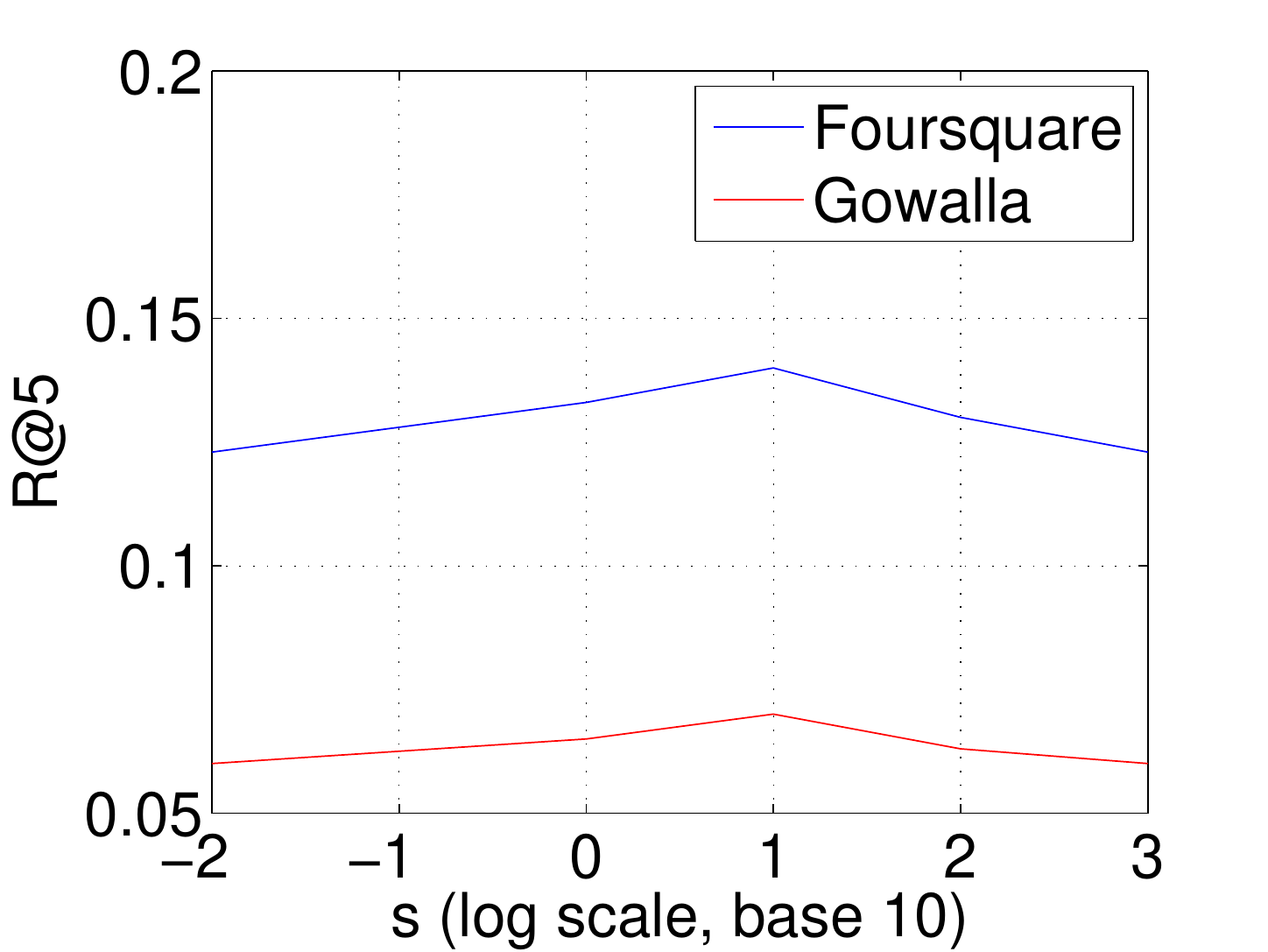}%
\label{subfig:s2}%
}
\vspace{-2mm}
\caption{Parameter effect on distance threshold $s$}
\label{fig:sparameter}

\end{figure}

In the \textit{GT-SEER} model, we classify the unchecked POIs as neighboring POIs and non-neighboring POIs to constitute a new preference set according to a threshold distance $s$. Here we choose different values of $s$ to see how this parameter affects the model performance, as shown in Figure~\ref{fig:sparameter} (we only show P@$5$ and R@$5$ for space limit). We observe that \textit{GT-SEER} model achieves the best performance at $s=10$. Furthermore, when $s$ is extremely small or extremely large, we cannot classify the unchecked POIs, hence the \textit{GT-SEER} model degenerates to \textit{T-SEER} model without the consideration of geographical influence. 

%
%
%


\section{Conclusion and Further Work}
\label{sec:cfw}
We study the problem of POI recommendation in this paper. 
In order to capture contextual check-in information hidden in the sequences, we propose the POI embedding model to learn POI representations. 
Next, we propose the \textit{SEER} model to recommend POIs, which learns user preferences via a pairwise ranking model under the sequential representation constraint modeled by the POI embeddings. 
Moreover, we establish the temporal POI embedding model to capture the temporal variance of sequences on different days and propose the \textit{T-SEER} model to incorporate this kind of temporal influence.
Finally, we propose  the \textit{GT-SEER} model to improve the recommendation performance through incorporating geographical influence into the \textit{T-SEER} model. 
Experimental results on two datasets, Foursquare and Gowalla, show that our sequential embedding rank model better captures the sequential pattern, outperforming previous sequential model \textit{LORE} more than 50\%.
In addition, the proposed \textit{GT-SEER} model improves at least 28\% on both datasets for all metrics compared with the best baseline competitor. 

Our future work may be carried out as follows: 1) Since we only consider the sequence of one day in this paper, we may discuss other scenarios in the future, for instance, sequences consisted of consecutive check-ins whose interval is under a fixed time threshold, e.g., four hours or eight hours. 
2) We may subsume more information, e.g., users' comments and social relations, in this system to improve performance.

\bibliographystyle{plain}
\bibliography{icdm}

\end{document}